\documentclass[aps,prl,twocolumn,amsmath,amssymb,nofootinbib,superscriptaddress,floatfix,reprint,longbibliography]{revtex4-2}
\usepackage{natbib}
\usepackage[T1]{fontenc}
\usepackage{amsmath}
\usepackage{amsfonts}
\usepackage{amsthm}
\usepackage{amssymb}
\usepackage{lipsum}
\usepackage{graphicx}
\usepackage[usenames,dvipsnames]{color}
\usepackage{float}
\usepackage{multirow}
\usepackage{soul}
\usepackage{appendix}
\usepackage[colorlinks=true, urlcolor=blue, linkcolor=blue, citecolor=blue, pdftex]{hyperref}
\usepackage{dcolumn}
\usepackage{bm}
\usepackage{url}
\usepackage{braket}
\usepackage[range-units=single,separate-uncertainty]{siunitx}
\usepackage{verbatim} 
\usepackage{booktabs}
\usepackage[version=4]{mhchem}

\usepackage[english]{babel}
\begin{document}
\author{Savita Chaudhary}
\affiliation{Department of Physical Sciences, Indian Institute of Science Education and Research (IISER) Mohali, Knowledge City, Sector 81, Mohali 140306, India.}

\author{Armando Consiglio}\affiliation{Institut f\"{u}r Theoretische Physik und Astrophysik and W\"{u}rzburg-Dresden Cluster of Excellence ct.qmat, Universit\"{a}t W\"{u}rzburg, 97074 W\"{u}rzburg, Germany}
\affiliation{CNR-IOM Istituto Officina dei Materiali, I-34139 Trieste, Italy}

\author{Jaskaran Singh}
\affiliation{Department of Physics, Punjabi University, Patiala 147002, India.}

\author{Domenico Di Sante}\affiliation{Department of Physics and Astronomy, Alma Mater Studiorum, University of Bologna, 40127 Bologna, Italy}

\author{Ronny Thomale}\affiliation{Institut f\"{u}r Theoretische Physik und Astrophysik and W\"{u}rzburg-Dresden Cluster of Excellence ct.qmat, Universit\"{a}t W\"{u}rzburg, 97074 W\"{u}rzburg, Germany}

\author{Yogesh Singh}
\affiliation{Department of Physical Sciences, Indian Institute of Science Education and Research (IISER) Mohali, Knowledge City, Sector 81, Mohali 140306, India.}

\date{\today}

\title{Tuning electronic correlations in the Kagome metals $RT_3$B$_2$}

\begin{abstract}

The $RT_3$B$_2$ ($R=$~Y, Lu, $T=$ Co, Os) family hosts a perfect kagome lattice of $T$ atoms, offering an interesting platform to investigate the interplay of electronic structure, superconductivity, and lattice dynamics. Here, we compare two members of this family, LuOs$_3$B$_2$ and YCo$_3$B$_2$, with similar crystallography but differing chemical composition, leading to distinct electronic correlation strengths and spin-orbit coupling effects. We confirm superconductivity in LuOs$_3$B$_2$ with $T_c = 4.75$K, while YCo$_3$B$_2$ remains non-superconducting above 1.8K. First-principles estimates of the electron-phonon coupling for LuOs$_3$B$_2$ are consistent with its observed $T_c$ and suggest a moderate coupling strength. Both materials exhibit kagome-derived electronic features, including quasi-flat bands, Dirac cones, and van Hove singularities. Fermi surface calculations reveal quasi-one-dimensional behavior along the $c$-axis in YCo$_3$B$_2$, in contrast to the more three-dimensional Fermiology of LuOs$_3$B$_2$. Phonon calculations for LuOs$_3$B$_2$ show imaginary modes, indicating potential lattice instabilities. Experimental estimates of the Wilson and Kadowaki-Woods ratios point to non-negligible electronic correlations in both compounds.
\end{abstract}

\maketitle 
\section{Introduction} 
The kagome lattice has attracted a lot of attention recently due to its ability to host novel physics because of its distinctive electronic structure \cite{Balents2010,Savary_2017,doi:10.1126/science.aay0668,Knolle:2018xhp}. Insulating materials with a kagome lattice host localized magnetic moments, serving as ideal systems to investigate geometrically frustrated magnetism. A prominent example is the quantum spin liquid (QSL) state observed in herbertsmithite where the kagome layers of copper ions exhibit long-range quantum entanglement and fractionalized spinon excitations without conventional magnetic order \cite{doi:10.1126/science.aab2120,Han2012}. Beyond two dimensions, insulating quantum magnets with three-dimensional kagome-inspired lattices have also demonstrated novel frustration-driven phenomena and QSL behavior \cite{Balz2016,PhysRevLett.99.137207,PhysRevB.88.220413}. Recent attention has shifted to metallic kagome systems, where the interplay of topology and electron correlations becomes possible. Theoretical predictions highlight the coexistence of Dirac cones, flat bands and van Hove singularities in the electronic band structure \cite{9e66d15ca73e41e8a8a2746de65fd232}, creating opportunities for exotic electronic states. Experimental advances have identified several families of metallic kagome materials, such as the predicted Herbertsmithite related material GaCu$_3$(OH)$_6$Cl$_2$ \cite{9e66d15ca73e41e8a8a2746de65fd232}, FeSn and CoSn compounds \cite{95810bfd2245410db75382c1581ffe9e,0e7cb127b339477390e474f0da9235e5}, and the 166 compounds e.g. YMn$_6$Sn$_6$ \cite{Li2021}, which realize these electronic features.  Recently, the $A$V$_3$Sb$_5$ ($A$ = K, Rb, Cs) family of materials has been discovered, featuring an ideal kagome lattice formed by vanadium ions. These compounds display signatures of strong electronic correlations and nontrivial topological characteristics, leading to novel behaviors like the emergence of charge density waves, superconductivity, and anomalous Hall effect \cite{PhysRevMaterials.3.094407,PhysRevLett.125.247002,PhysRevMaterials.5.034801,doi:10.1126/sciadv.abb6003}.

$RT_3X_2$ ($R$ = lanthanide, $T$ = 3$d$, 4$d$ or 5$d$ transition metal, $X$ = Si, B) is another family of materials hosting the kagome lattice which was discovered in the 1980s and 1990s \cite{KU198091,BARZ19801489,VANDENBERG19801493}, and several of its members were found to be superconducting with T$_c$ between 1 and $\approx 7 K$ \cite{KU198091,BARZ19801489,VANDENBERG19801493,PhysRevB.32.4742,ATHREYA1985330,PhysRevB.30.444}. The majority of these investigations, however, did not make a connection of physical properties with the underlying kagome lattice.  Recently, several unconventional properties have been reported in LaRu$_3$Si$_2$, the material having the highest $T_c = 7$~K in this family, ThRu$_3$Si$_2$ and YRu$_3$Si$_2$, possibly arising from electron correlations from the flat bands \cite{PhysRevB.84.214527,PhysRevB.86.024513,PhysRevB.94.094523,PhysRevMaterials.5.034803,Liu_2024}. We recently reported on LaRh$_3$B$_2$, which contains the essential fermiology of the 2D kagome lattice.  The material shows superconductivity at $T_c \sim 2.6$~K but did not show any experimental evidence of the charge density wave (CDW) state.  Lack of significant electronic correlations in LaRh$_3$B$_2$ suggests that electronic correlations might play an important role in the existence of CDW states observed in other Kagome metals \cite{PhysRevB.107.085103}. The $RT_3X_2$ family, which can be synthesized for $T =$ 3$d$, 4$d$, or 5$d$ transition metals, offers the opportunity to tune electronic correlations and spin-orbit coupling (SOC). 

In this work, we have synthesized and studied $RT_3$B$_2$ ($R =$ Y, Lu, and $T =$ Co, Os) through measurements of their electronic transport, magnetic susceptibility, and heat capacity and through calculations of their electronic structure, Fermi surface, and phonon spectrum.  Our measurements show that YCo$_3$B$_2$ and to a greater extent LuOs$_3$B$_2$ host strong electronic correlation, as evidenced by the enhanced values of the Wilson and the Kadowaki-Woods ratios. Additionally, the LuOs$_3$B$_2$ phonon spectra show imaginary modes pointing to a susceptibility to lattice instabilities.  Our work therefore provides a connection between electron correlation and structural or CDW-like tendencies in $RT_3$B$_2$ Kagome metals.   

\begin{figure*}[t]   
\includegraphics[width=5 in]{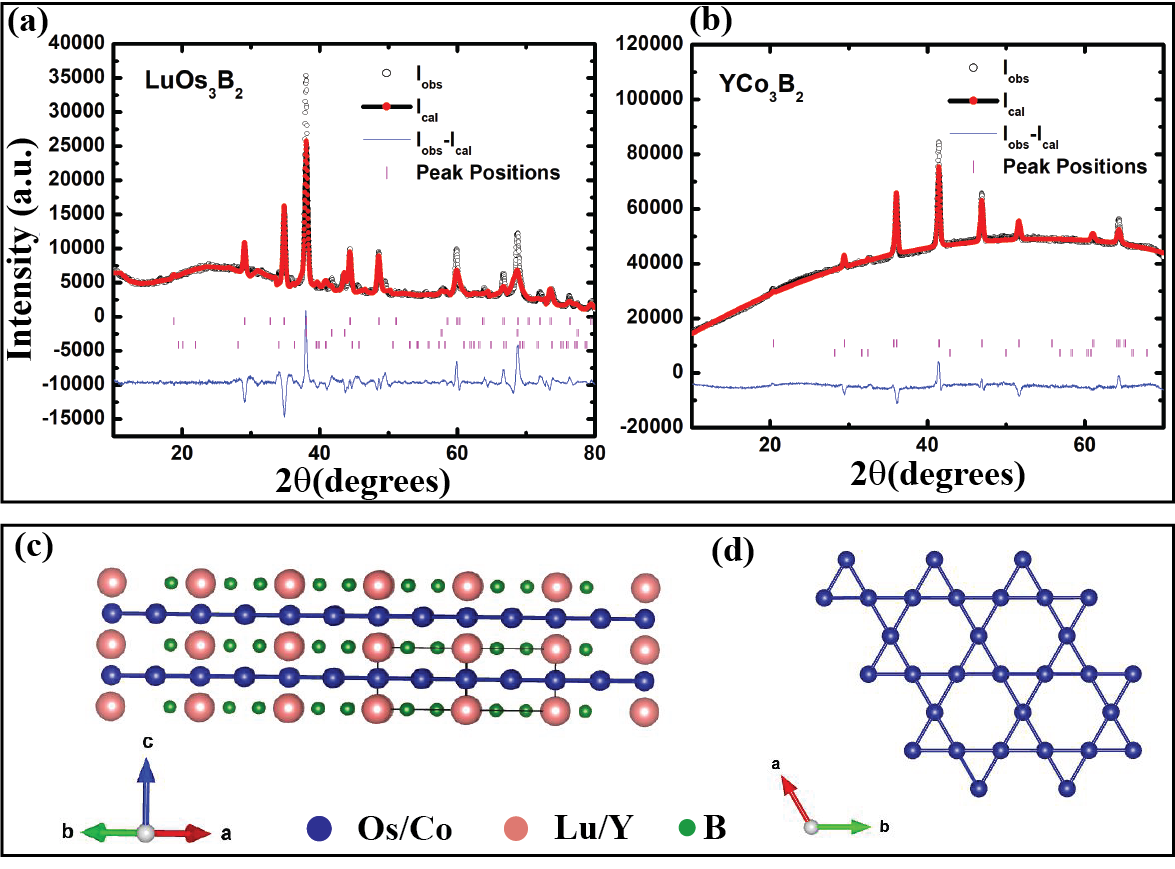}
\caption{(Color online) Powder x-ray diffraction and results of refinement for (a) LuOs$_3$B$_2$ and (b) YCo$_3$B$_2$. A schematic of the crystal structure of $RT_3$B$_2$ viewed (c) perpendicular to the crystallographic $c$-axis showing the layered nature of the structure with $T$ atomic planes separated along the $c$-axis by planes made up of $R$ and B atoms.  (d) Viewed along the $c$-axis, the $T$ atoms form an undistorted kagome lattice. 
\label{Fig-xrd}}
\end{figure*}

\section{Methods}
Polycrystalline samples of $RT_3$B$_2$ ($R =$ Y, Lu, $T =$ Co, Os) were synthesized by arc-melting stoichiometric amounts of high purity elements. The melted buttons were flipped over and melted $5$--$10$ times to promote homogeneity.  Powder X-ray diffraction (PXRD) on a Bruker D8 Advance diffractometer system with Cu-K$\alpha$ radiation was used to determine the phase purity of the synthesized materials. The relative stoichiometry of $R$ and $T$ was confirmed using energy dispersive spectroscopy using a scanning electron microscope. The dc magnetic susceptibility $\chi$, heat capacity $C$, and electrical transport were measured using a Quantum Design Physical Property Measurement System equipped with a He3 insert.   To theoretically simulate the electronic structure of $RT_3$B$_2$, we performed first-principles density functional theory (DFT) calculations using the Vienna Ab initio simulation package (VASP) \cite{PhysRevB.47.558,PhysRevB.49.14251,PhysRevB.54.11169,KRESSE199615}. We considered the projector-augmented wave (PAW) pseudo potential with exchange-correlation functional of generalized gradient approximation (GGA) of Perdew-Burke-Ernzerhof \cite{PhysRevB.59.1758,PhysRevLett.77.3865}. We adopted a $9\times9\times12$ $k$-mesh for the first Brillouin zone, and an energy cut-off of $550$~eV for the plane wave basis. The convergence criteria for energy is set to $10^{-6}$~eV. When considered, SOC was included self-consistently.  Phonon calculations have been performed using density functional perturbation theory, as implemented in Quantum Espresso \cite{10.1063/5.0005082,Giannozzi_2009,Giannozzi_2017}. 
 Exchange and correlation effects were included using the generalized gradient approximation (GGA) with the Perdew-Burke-Ernzerhof (PBE) functional \cite{PhysRevLett.77.3865}; the pseudopotentials are norm-conserving, with core correction, and scalar relativistic \cite{PhysRevB.88.085117}.\\
Self-consistent calculations of the previously relaxed unit cell have been performed with a 6$\times$6$\times$12 $k$-grid. The kinetic energy cutoff for the wavefunctions is equal to 100 Ry, while the cutoff for charge density is 400 Ry.  Convergence threshold for ionic minimization and electronic self-consistency are set to be 1.0D-15.  The self-consistency threshold for phonon calculations is 1.0D-15 as well, with a $q$-grid of 3$\times$3$\times$6.  Non-self consistent calculations for the density of states have been performed with a 30$\times$30$\times$30 grid.  Finally, the electron-phonon interaction is computed via an interpolation over the Brillouin Zone \cite{wierzbowska2006originslowhighpressurediscontinuities}.

\begin{figure*}[t]   
\includegraphics[width = 0.98\textwidth, trim={2.5cm 66cm 2.0cm 3.0cm},clip]{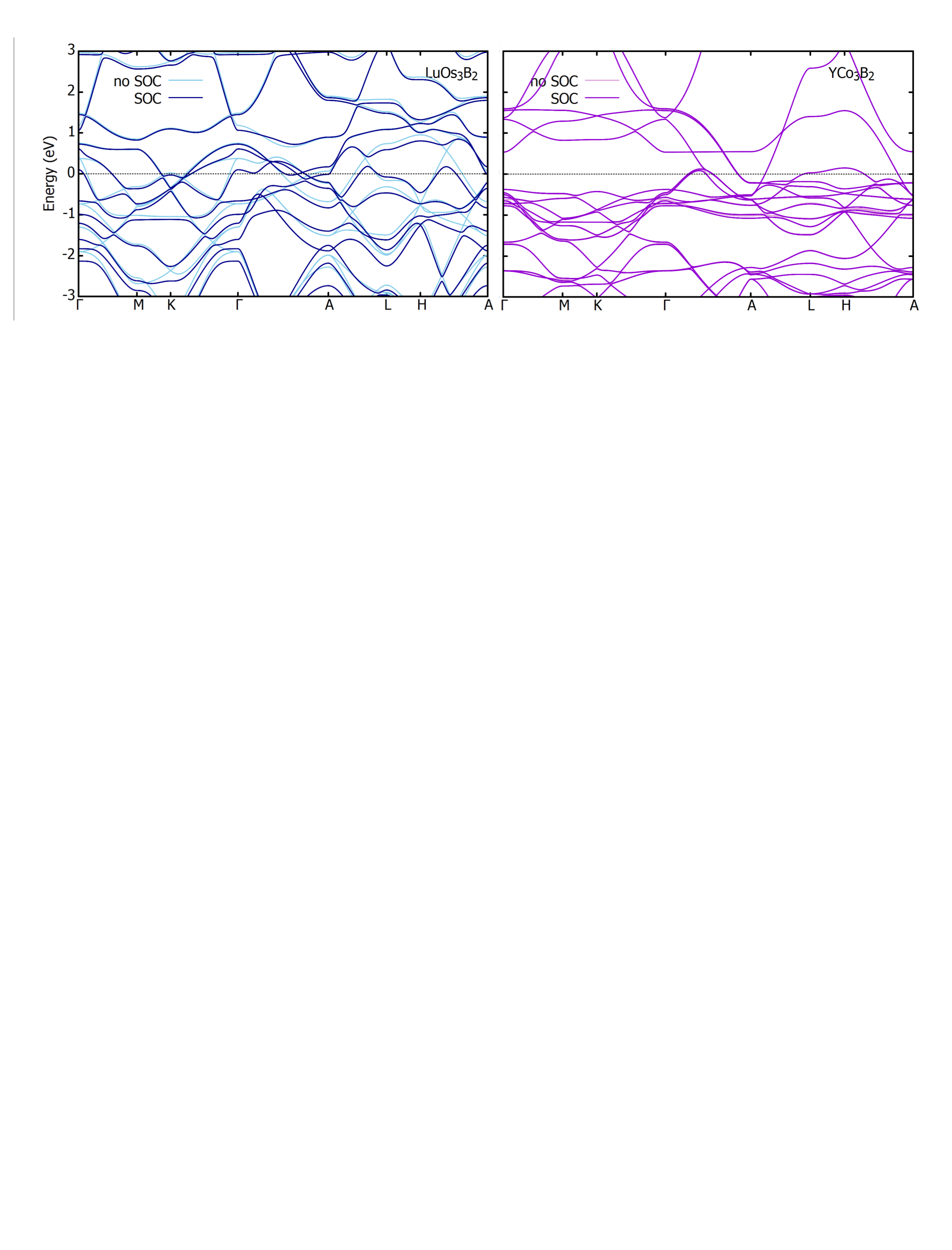}    
\caption{Electronic structure along high-symmetry paths in $k$-space, comparing with and without SOC cases, for LuOs$_3$B$_2$ (left panel) and YCo$_3$B$_2$ (right panel).  
\label{Fig-dos}}
\end{figure*}

\begin{figure*}[t]   
\includegraphics[width = 0.98\textwidth, trim={2.5cm 69cm 4.0cm 1.0cm},clip]{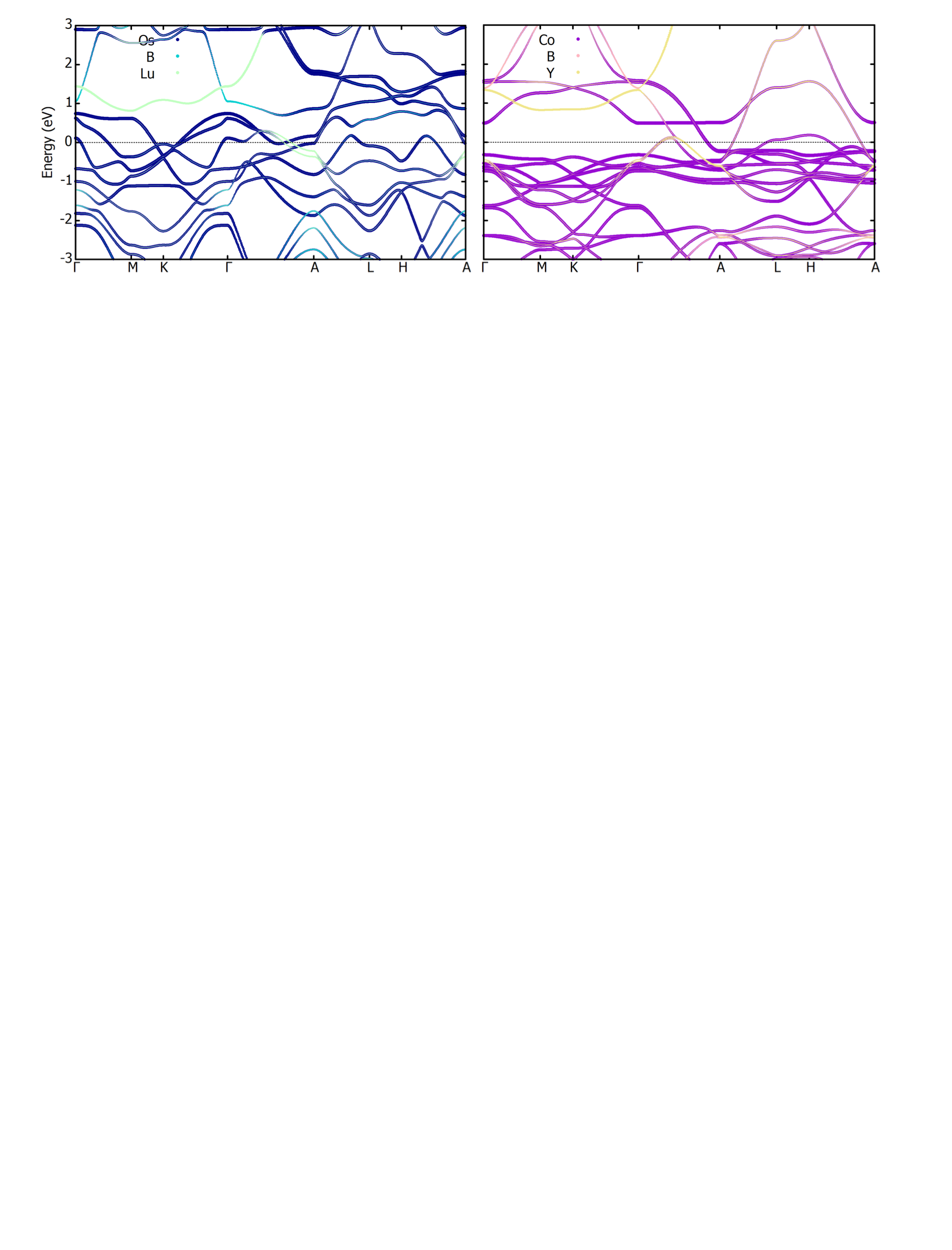}    
\caption{Elemental contributions to the electronic band structure of LuOs$_3$B$_2$ (left) and YCo$_3$B$_2$ (right). 
\label{Fig-elemental-dos}}
\end{figure*}


\section{Structure}
Both samples, LuOs$_3$B$_2$ and YCo$_3$B$_2$ crystallize in a honeycomb structure with space group P6/$mmm$.  With the exception of the unit cell size, the structure has no variable parameters.  A Rietveld refinement of the powder patterns  on the basis of hexagonal, P6/$mmm$ symmetry as shown in Fig.~\ref{Fig-xrd}(a) and Fig.~\ref{Fig-xrd}(b); gave the lattice parameters $a = 5.461$~\AA, $c = 3.070$~\AA; $a = 5.031$~\AA, $c = 3.036$~\AA~ for LuOs$_3$B$_2$ and YCo$_3$B$_2$, respectively.  These lattice parameters are in good agreement with prior reports \cite{Chacon2001,KU198091}.  From the powder XRD data we find that there are two additional phases i.e. $\approx 5$\% Lu$Os_2$, and $\approx 4$\% Os in case of LuOs$_3$B$_2$.  These impurity phases did not allow a good Rietveld refinement.  Therefore, only lattice parameters from the positions of the Bragg peaks will be quoted and no other refinement parameters are used in our study.  Repeated attempts with varying starting stoichiometry to produce single phase samples were unsuccessful.  This indicates that LuOs$_3$B$_2$ is most likely incongruently melting in nature and can't be produced as single phase from the melt. We also observed a tiny amount of Y ($\approx 1$\%) in the XRD data for YCo$_3$B$_2$. 

Fig.~\ref{Fig-xrd} (d) shows the schematic of the crystal structure of the samples, where the central atoms Os and Co are arranged in perfect kagome layers. These kagome layers of Os and Co are seperated by the respective layers of Lu and B, and Y and B are piled up along the $c$-axis as shown in Fig.~\ref{Fig-xrd} (c). Thus, the description of the crystal structure hints that these substances possess the structural components necessary to exhibit the electronic structure characteristics anticipated of a Kagome metal.  
It must be noted however, that the short $c$-axis could imply that coupling between kagome planes may be significant.

\section{Results}
\subsection{Electronic Structure}



Figure~\ref{Fig-dos} shows the electronic band structure with and without SOC for LuOs$_3$B$_2$ and YCo$_3$B$_2$ along some high-symmetry directions in the Brillouin zone.  The metallic nature of both materials is evident as several bands can be seen crossing the Fermi level.  The first thing that is evident is that electronic correlations play a larger role in YCo$_3$B$_2$ while SOC plays a larger role in LuOs$_3$B$_2$.  Indeed, the electronic bands in YCo$_3$B$_2$ are less dispersive and narrow compared to LuOs$_3$B$_2$, indicating enhanced correlations in YCo$_3$B$_2$.  Whereas the inclusion of SOC has the most significant effect in the band structure of LuOs$_3$B$_2$ and is almost negligible for YCo$_3$B$_2$.  We also note the high density of states in YCo$_3$B$_2$ at energies between $0.5$ and $1$~eV below the Fermi energy $E_F$.  It could be of interest to enhance the density of states at $E_F$ by shifting the valence bands upward via strain/pressure or doping.  

We now turn to the features of the band structure expected to arise from the kagome lattice.  The band structure of both materials show a quasi-flat band (FB) in the $\Gamma-M-K-\Gamma$ direction.  YCo$_3$B$_2$ shows additional flat-band features in the $\Gamma - A$ direction.  Dirac cones (DC) are observed at the $K$ point of the Brillouin zone.  The DCs are closer to $E_F$ for LuOs$_3$B$_2$. We also identify van Hove (VH) singularities located at the $M$ points.  

The kagome lattice related features in the band structure arise primarily from the Co and Os orbitals as shown in the element specific band structures shown in Fig.~\ref{Fig-elemental-dos}. It is also evident that bands close to $E_F$ mostly arise from the Kagome network.  Figure~\ref{Fig-elemental-dos} shows the orbital contributions from Co and Os to the band structure for both materials.  We note that for YCo$_3$B$_2$ there are no states at $E_F$ for $k_z = 0.0$.  Most of the states at $E_F$ come from $d_{xz}$ and $d_{yz}$ orbitals, at $k_z = 0.5$.  For LuOs$_3$B$_2$, $d_{xz}$ and $d_{yz}$ orbitals contribute more for $k_z = 0.0$ while $d_{x^2-y^2}$ and $d_{xy}$ contribute more for $k_z = 0.5$.  

We recall that we previously reported similar kagome band-structure features for LaRh$_3$B$_2$ \cite{PhysRevB.107.085103} indicating that the $RT_3$B$_2$ family of kagome metals possesses the predicted features of the 2D kagome lattice band structure near $E_F$ with modifications arising most likely from the three-dimensional nature of the material.  


\subsection{Physical Properties}

\begin{figure*}[t]   
\includegraphics[width = 6.5in]{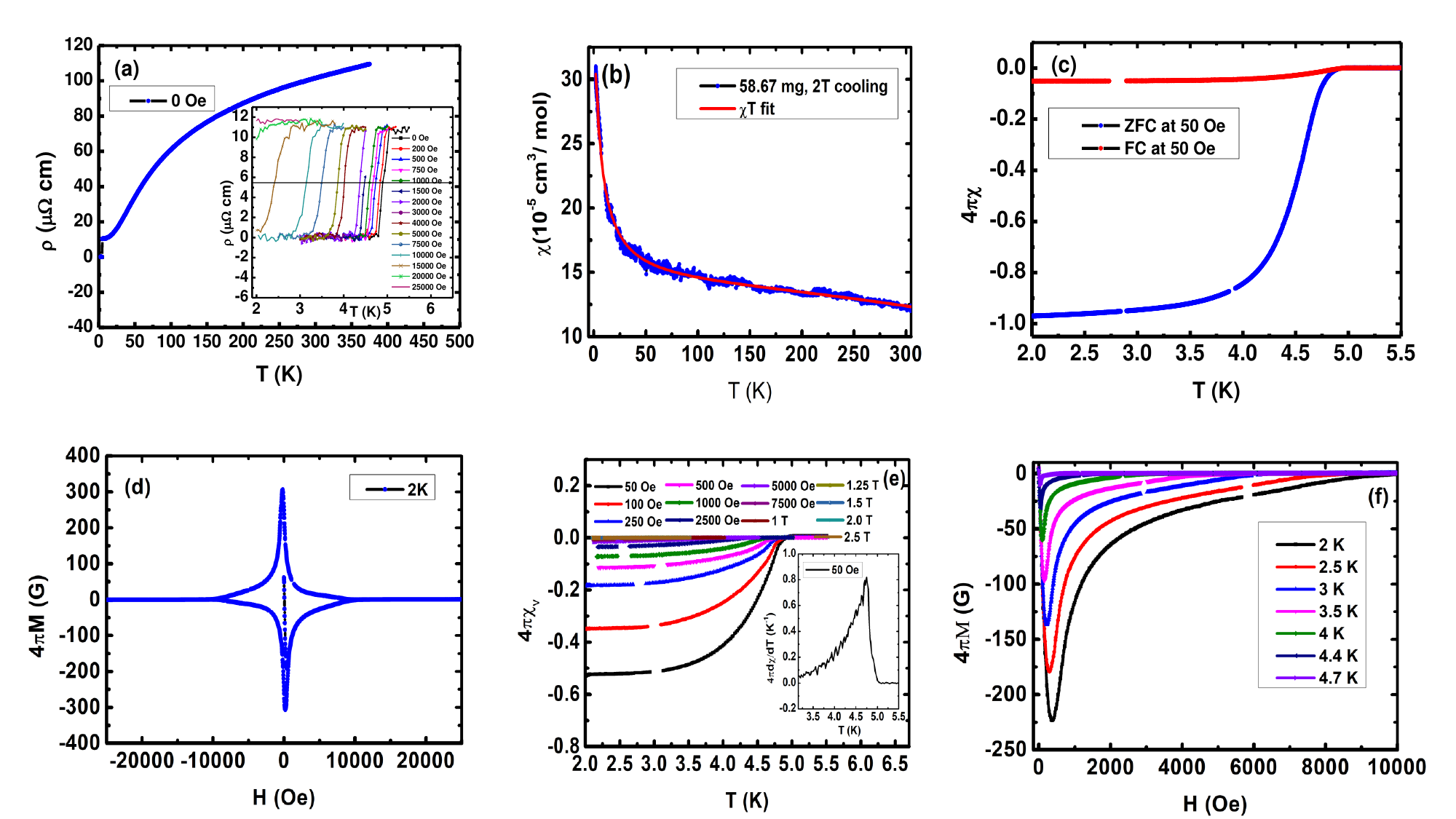}    
\caption{(Color online) (a) shows the electrical resistivity of LuOs$_3$B$_2$ measured in the temperature range of 2.5 to 375 K at zero field, the inset shows the resistivity data between 4 to 5.5 K at various applied fields, (b) shows the magnetic susceptibility $\chi$ versus temperature T curve plotted between 2 and 305 K in an applied magnetic field of H = 2 T, (c) shows the temperature dependence of the dimensionless magnetic susceptibility 4$\pi$$\chi$ in the zero field cooled warming and the field cooled cooling mode measured at 50 Oe, (d) shows the hysteresis loop at 2 K of the volume magnetization $M_v$(H) normalized by 1/4$\pi$ , versus applied magnetic field H, (e) shows the temperature dependence of the field cooled dimensionless magnetic susceptibility (4$\pi$$\chi$) at various fields and the inset shows the derivative of 4$\pi$$\chi$ w.r.t T, (f) shows the volume magnetization M$_v$(H) normalized by 1/4$\pi$ versus applied magnetic field H at various temperatures.}
\label{Fig-2}
\end{figure*}

We now discuss the electrical, magnetic, and thermal properties of LuOs$_3$B$_2$ in the normal and superconducting states. Fig~\ref{Fig-2} (a) shows the electrical resistivity of LuOs$_3$B$_2$ measured in the temperature range of $2$ to $375$~K at zero field, which confirms the metallic behavior of LuOs$_3$B$_2$ with a residual resistivity ratio $\mathrm{RRR}=\rho(375 \mathrm{~K}) / \rho(5 \mathrm{~K}) \approx 11$. 
The $\rho(T)$ in the low temperature region was fit by the equation $\rho(T)=\rho_0(T)+aT^2+bT^5$, where $\rho_0(T)$ is the residual resistivity and the terms a$T^2$ and b$T^5$ arise from electron–electron and electron–phonon scattering, respectively.  
Fitting the $\rho(T)$ data in the temperature range (5 K < T < 35 K) gave the fit parameters $\rho_0 = 9.68 \mu\Omega cm$, $a = 9.7 \times 10^{-3} \mu\Omega$~cm~K$^{-2}$, and $b = 3.03 \times 10^{-8} \mu\Omega$~cm~K$^{-5}$.  Considerably small value of b as compared to a for $T < 35$~K suggests that the resistivity in low temperature region is mainly governed by electron–electron scattering.  The inset shows the resistivity data from $T = 4$ to $5.5$~K at various applied fields to highlight the onset of superconductivity in LuOs$_3$B$_2$ indicated  by the sharp drop to zero resistance below $T_c \approx$  4.9 K in zero field.  This is consistent with $T_c$ values reported previously \cite{KU198091,PhysRevB.32.4742}.  The suppression of the superconducting transition to lower temperature on increasing magnetic field is evident in the resistivity data shown in the inset of Fig~\ref{Fig-2} (a).  Figure~\ref{Fig-2} (b) shows the magnetic susceptibility $\chi$ versus temperature $T$ curve plotted between $2$ and $305~$K in an applied magnetic field of $H = 2$~T.  For a Pauli paramagnetic metal, a $T$ independent magnetic susceptibility is expected, described by: $\chi_p=\mu_B^2 D\left(\varepsilon_F\right)$, where $\mu_B$ is the Bohr magneton and $D\left(\varepsilon_F\right)$ is the electronic density of states at the Fermi level.  However, in LuOs$_3$B$_2$, the susceptibility is found to be temperature-dependent across the entire measured temperature range.  This observation is similar to LaRh$_3$B$_2$ which we reported previously \cite{PhysRevB.107.085103}.  The Curie-like upturn at low temperatures in $\chi(T)$ is expected to arise from trace amounts of magnetic impurities in the starting elements used to synthesize the materials.  We have therefore fitted the $\chi(T)$ in the full temperature range with the expression $\chi(T)=\chi_0\left[1-\left(T / T_E\right)^2\right]+C /(T-\theta)$, where the first term represents the $T$-dependent Pauli paramagnetic susceptibility, the second term accounts for contributions from trace amounts of magnetic impurities.  The fitting parameters are the temperature-independent average Pauli paramagnetic susceptibility, $\chi_0$; a phenomenological parameter related to the Fermi energy, $T_E$; the Curie constant of the impurities, C; and the Weiss temperature $\theta$, representing any interactions between the magnetic impurities. 
The fitting curve shown as the solid curve through the data in Fig~\ref{Fig-2}~(b) estimates $\chi_0 = 13.4(1) \times 10^{-5}$~cm$^3$/mol, T$_E = 884(7)$~K, $C = 0.00138(1)$~cm$^3$~K/mol, and $\theta = -6.1(1)$~K\@.  The obtained value of $C$ is equivalent to 0.35\% of S = 1/2 impurities, which is quite small.  

The temperature independent susceptibility can be expressed as $\chi_0 = \chi_{core} + \chi_P + \chi_{VV} + \chi_L$ where $\chi_{core}$ is the diamagnetic orbital contribution from the electrons (ionic or atomic), $\chi_P$ is the Pauli paramagnetic susceptibility of conduction electrons, $\chi_{VV}$ is Van Vleck paramagnetic orbital contribution and $\chi_L$ is the Landau orbital diamagnetism of conduction electrons ($\chi_L \approx \frac{-1}{3} \chi_P$). Assuming covalent nature of bonds in LuOs$_3$B$_2$, we have taken the atomic diamagnetism values $\chi_{core}$ \cite{PhysRevA.2.1130} for Lu (-60.55 × 10$^{-6}$ cm$^3$/mol), Os (-53.82 × 10$^{-6}$ cm$^3$/mol) and B (-12.54 × 10$^{-6}$ cm$^3$/mol) from which we get $\chi_{core}$ = -2.47 × 10$^{-4}$ cm$^3$/mol. The Van Vleck paramagnetic orbital contribution can be neglected giving $\chi_P = 5.72 \times 10^{-4}$~cm$^3$/mol.

We now present observation of superconductivity.  Figure~\ref{Fig-2} (c) shows the temperature dependence of the dimensionless magnetic susceptibility 4$\pi$$\chi$ in the zero field cooled (ZFC) warming and the field cooled (FC) cooling mode measured in a magnetic field of $50$~Oe.
The presence of a diamagnetic signal in the ZFC and the FC data below $T_c \approx 4.75$~K confirms the superconducting transition in LuOs$_3$B$_2$. The field cooled data has a weak diamagnetic signal which points towards strong pinning in the sample suggesting type II superconductivity in LuOs$_3$B$_2$ which we confirm below using magnetization measurements. Figure~\ref{Fig-2} (d) shows the hysteresis loop at 2 K of the volume magnetization $M_v$(H) normalized by 1/4$\pi$, versus applied magnetic field H. The hysteresis loop is typical of a Type-II superconductor.  Figure~\ref{Fig-2}~(e) and ~(f) show the temperature dependence of 4$\pi$$\chi$ at various fields and $4\pi$M$_v$(H) at various temperatures respectively. We will use these data to extract the upper critical field as a function of temperature. 

\begin{figure*}[t]
\centering
\includegraphics[width=6.8 in]{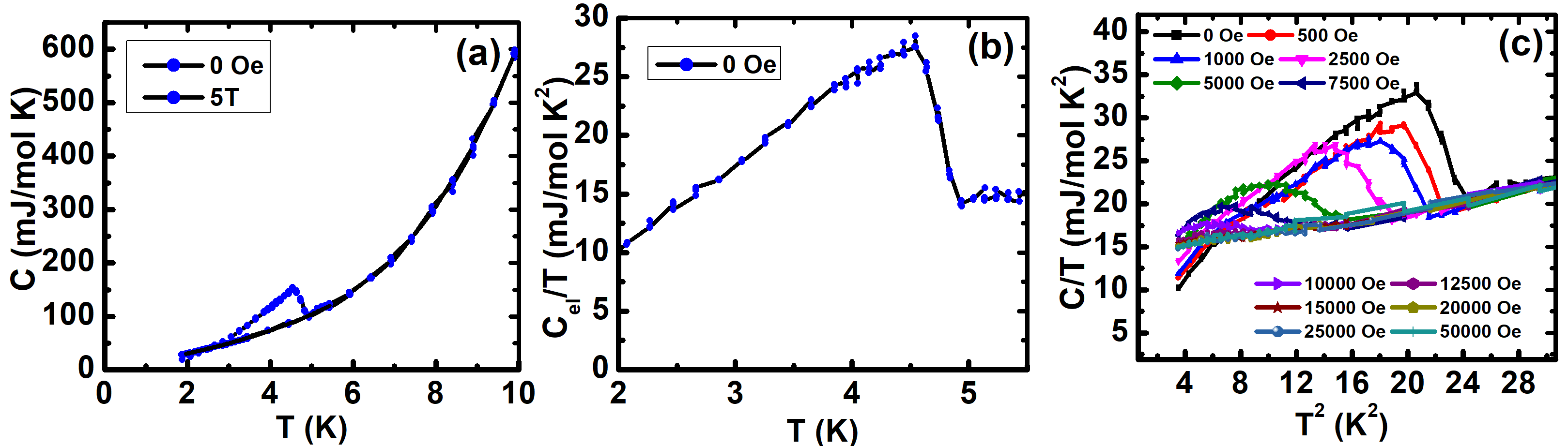}
\caption{(a) Heat capacity between 2 to 10 K at 0 Oe as well as at 5~T, (b) the electronic heat capacity C$_{el}$ divided by T versus T and (c) $C/T$ versus T$^2$ at various magnetic fields.}
\label{2}    
\end{figure*}

The heat capacity measurements provide evidence for the bulk nature of superconductivity in LuOs$_3$B$_2$.  Figure~\ref{2} (a) shows the heat capacity $C$ measured between $2$ and $10$~K at $0$~Oe as well as at $5$~T\@.  The anomaly with an onset at $\approx 4.9$~K signals a transition into the superconducting state.  The $C(T)$ data in $5$~T then provides the normal state heat capacity.  The heat capacity in the normal state can be expressed as: $C=\gamma_n T+\beta T^3$ where $\gamma_n$ is the normal state Sommerfeld coefficient and the second term accounts for the lattice contribution. A fit of the normal state heat capacity data to the above expression gave the fitting parameters $\gamma_n = 14.28(9)$~mJ/mol$K^2$ and $\beta = 0.261(6)$~mJ/mol$K^4$. Using this value of $\beta$, we can calculate the Debye temperature of $\theta_D = 355(3)$~K \cite{PhysRevB.76.214510}. The electronic heat capacity can then be extracted from total heat capacity by subtracting the lattice contribution.  Fig~\ref{2} (b) shows the resultant electronic heat capacity C$_{el}$ divided by T versus T, which reveals a transition from the normal to the superconducting state with critical temperature $T_c \approx 4.75$~K. Initial analysis of the heat capacity data measured down to 2 K hints towards the possibility of multi-gap superconductivity in LuOs$_3$B$_2$ because the normalized heat capacity jump at T$_c$ of $\frac{\Delta C}{\gamma_nT_c} \approx 1$ magnitude is smaller than the value expected ($1.43$) for a single gap BCS superconductor.  Figure~\ref{2} (c) shows $C/T$ versus T$^2$ at various magnetic fields showing the suppression of the critical temperature on the application of a magnetic field. We will use this data to extract the upper critical field as a function of temperature.

We can now calculate the Wilson ratio ($R_W$) to get some idea about the strength of electronic correlations in LuOs$_3$B$_2$ \cite{RevModPhys.47.773}. The Wilson ratio is defined as the ratio of density of states obtained from the magnetic measurements to the density of states obtained from the heat capacity measurements i.e. $R_W = \frac{\pi^2 K_B^2}{3 \mu_B^2} \frac{\chi_P}{\gamma}$, $R_W$ = 1 for a free-electron Fermi gas. Substituting $\chi_P$ = 57.2(1) × 10$^{-5}$ cm$^3$/mol and $\gamma_n = 14.28(9)$~mJ/mol K$^2$ for LuOs$_3$B$_2$, we get $R_W \approx 3$ at $T = 0$~K. The enhanced value of the Wilson ratio indicates strong correlations in the material, similar values of $R_W$ have earlier been reported for kagome metals LaRu$_3$Si$_2$ \cite{PhysRevB.84.214527}, ThRu$_3$Si$_2$ \cite{Liu_2024}, and YRu$_3$Si$_2$ \cite{Gong_2022}.  

To further support the strength of electronic correlation in LuOs$_3$B$_2$ as suggested by the Wilson Ratio, we now calculate the Kadowaki–Woods ratio to \cite{PhysRevLett.20.1439,KADOWAKI1986507}, which is defined as $A/\gamma^2$, where A is the coefficient of the quadratic term in the low temperature resistivity that occurs due to electron-electron scattering and $\gamma$ is the normal state Sommerfeld coefficient in the heat capacity. Substituting $A = 9.7  \times 10^{-3} \mu\Omega$ cm $K^{-2}$ and $\gamma$ = 14.28 mJ/mol $K^2$, we get KWR $\approx 48 \mu\Omega$ cm mol$^2$ K$^{2}$~J$^{-2}$ which indicates significant electronic correlations in LuOs$_3$B$_2$.

\begin{figure}
\centering
\includegraphics[width=8.5cm]{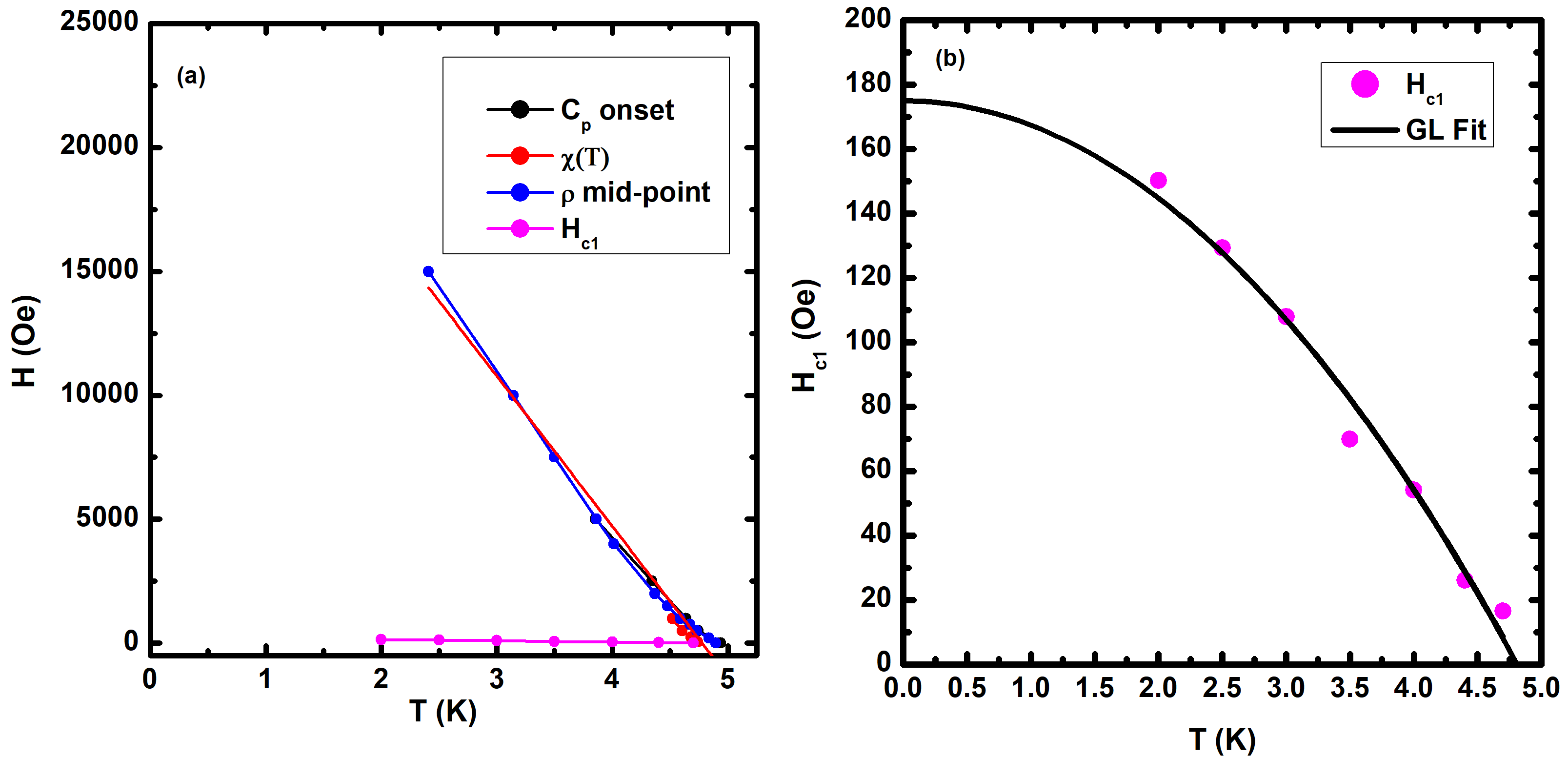}
\caption{(a) Variation of the lower critical field $H_{c1}$ and the upper critical field $H_{c2}$ as a function of temperature T  from various measurements in finite magnetic field and (b) fitting of the lower critical field data to the Ginzburg Landau equation.}\label{3}
\end{figure}

We now estimate various superconducting parameters using expressions mentioned in references \cite{PhysRevB.76.214510,PhysRevB.82.144532}.  We first extract the variation of the lower and upper critical fields $H_{c1}$ and $H_{c2}$ as a function of the temperature $T$ from various field-dependent measurements presented above. The critical field data so obtained are shown in Fig~\ref{3}~(a).  All the data points of upper critical feild $H_{c2}$ from different measurements are in excellent agreement with each other but we observe an unusual quasi-linear dependence in the H-T phase diagram with the slope $\left.\frac{d H_{c 2}}{d T}\right|_{T_c}= -6055 \text { Oe/K }$  which is highly unusual. The Werthamer-Helfand-Hohenberg (WHH) formula \cite{PhysRevB.82.144532} for the clean limit is used to get an estimate of $H_{c2}(0)$ given by the expression $H_{c 2}(0)=-\left.0.693 T_c \frac{d H_{c 2}}{dT}\right|_{T_c}=20.6 \text { K Oe }$. This value of $H_{c2}(0)$ is then used to estimate the coherence length $\xi$ given by the expression $H_{c 2}=\phi_0 / 2 \pi \xi^2$, where $\phi_0= hc/2e = 2.068 \times 10^{-7} G cm^2$ is the flux quantum. We can now estimate $\xi = 12.57 nm$, using H$_{c2}(0)$ = 20.6 KOe and $T_c\approx 4.9$ K. The H$_{c1}$ data is well fitted with the Ginzburg-Landau expressions: $H_{c1}(T)= H_{c1}(0)[1-[\frac{T}{T_c}]^2]$ where $H_{c1}(0)$ is the lower critical field at 0K and $T_c$ is the transition temperature. The fitting shown in Fig~\ref{3}~(b) gave the parameters H$_{c1}(0) = 175$~Oe and $T_c = 4.82$~K\@. The transition temperature obtained from the GL fit is in good agreement with bulk measurements presented previously. We can use the above values of $\xi$ and $H_{c1}(0)$ to evaluate the penetration depth $\lambda$ and Ginzburg-Landau parameter $\kappa$ using the relation : $\mu_0 H_{c 1}(0)=\frac{\ln (\lambda / \xi) \Phi_0}{4 \pi \lambda^2}$ and $\kappa = \lambda/\xi$, which gives $\lambda = 153.2$~nm and $\kappa = 12.2$. The large value of $\kappa$ indicates that LuOs$_3$B$_2$ is a Type-II superconductor. Using above values of $H_{c1}(0)$, $H_{c2}(0)$ and $\kappa$; we can also estimate the thermodynamic critical field $\mu_0H_c(0)$ at 0 K by the expression: $H_{c1}(0) \cdot H_{c2}(0) = H^2_{c}(0)$ln$(\kappa)$; which gives $H_{c}(0) = 1.2$ T. All superconducting parameters are collected in Table~\ref{T_TC}.

From $T_c$, an estimate of the electron-phonon coupling constant $\lambda_{ep}$ can be made using McMillan’s formula, which relates $T_c$ to $\lambda_{ep}$, the Debye temperature $\theta_D$, and the Coulomb repulsion constant $\mu^*$ \cite{PhysRevB.82.144532}
\\
$$
T_c=\frac{\theta_D}{1.45} \exp \left[-\frac{1.04\left(1+\lambda_{\mathrm{ep}}\right)}{\lambda_{\mathrm{ep}}-\mu^*\left(1+0.62 \lambda_{\mathrm{ep}}\right)}\right]
$$
which can be inverted to give $\lambda_{ep}$ in terms of T$_c$, $\theta_D$ and $\mu^*$ as
$$
\lambda_{\text {ep }}=\frac{1.04+\mu^* \ln \left(\frac{\theta_D}{1.45 T_c}\right)}{\left(1-0.62 \mu^{*}\right) \ln \left(\frac{\theta_D}{1.45 T_c}\right)-1.04}
$$
We obtain $\lambda_{ep}$ = 0.54 and 0.64 for $\mu^*$ = 0.10 and 0.15, respectively by using $\theta_D$ = 355 K and using T$_c = 4.75$~K\@. These values of $\lambda_{ep}$ suggest moderate electron-phonon coupling in LuOs$_3$B$_2$.

\begin{table}
\begin{center}
\caption{Normal and superconducting state parameters for  LuOs$_3$B$_2$. Here $\gamma$ is the Sommerfeld coefficient, $\beta$ is the coefficient of the $T^3$ term in the low temperature heat capacity, $\theta_{\rm D}$ is the Debye temperature, $\xi$ is the superconducting coherence length, $\lambda$ is the penetration depth.}
\vspace{0.5cm}
\label{T_TC}
\begin{tabular}{|c|c|}
\hline
\hline
RRR& $\approx 11$ \\ \hline
$\gamma$~(mJ/mol~K$^2$) & 14.28 \\ \hline
$\beta$~(mJ/mol~K$^4$) & 0.261 \\ \hline
$\theta_{\rm D}$~(K) & 355 \\ \hline
{H$_{c1}(0)$(Oe)} & {175} \\ \hline
{H$_{c2}(0)$(K Oe)} & {20.6} \\ \hline
{H$_{c}(0)$(T)} & {1.2} \\ \hline
{T$_{\rm C}$(K)} & {4.9} \\ \hline
$\xi_{GL}$(nm)& 12.57\\ \hline
$\lambda_{GL}$(nm) & 153.2 \\ \hline
$\kappa_{GL}$ & 12.2 \\ \hline
$\lambda_{ep}$ & 0.54 \\ \hline

\hline 
\end{tabular}
\end{center}
\end{table}    

\begin{figure*}[b]   
\includegraphics[width = 7in]{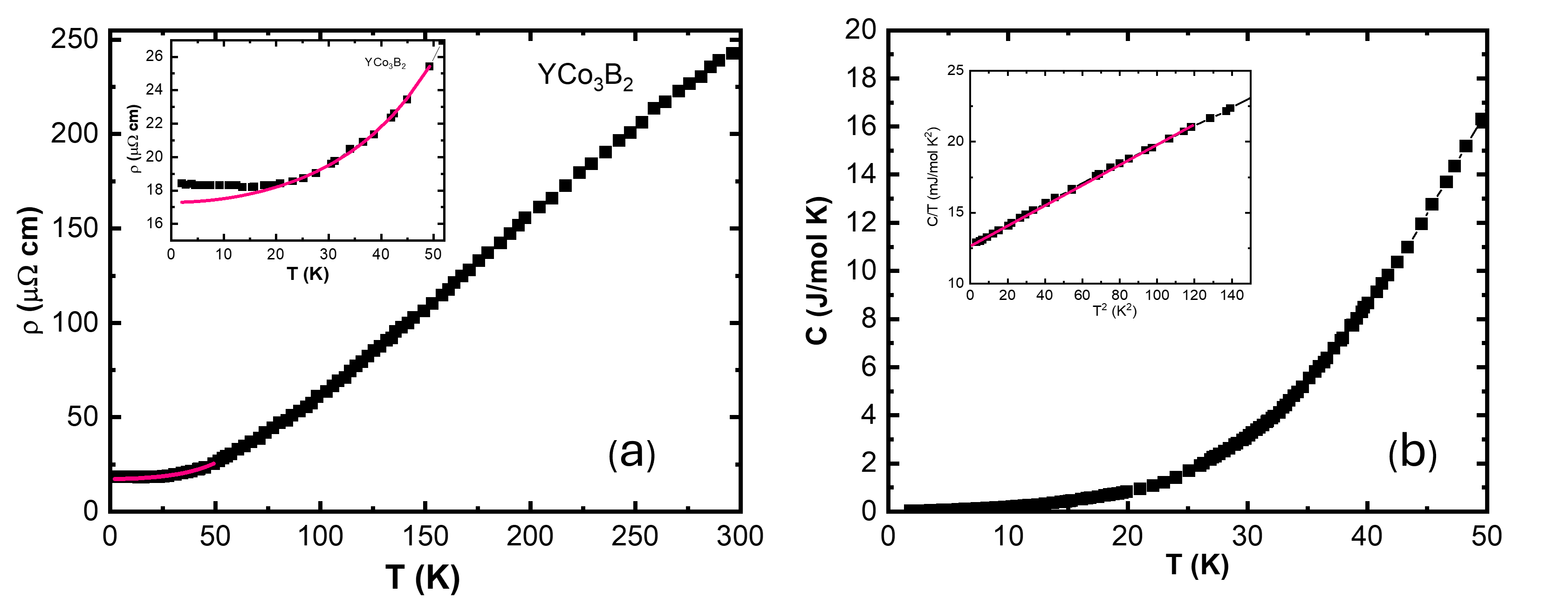}  
\caption{(Color online) (a) Resistivity and (b) Heat Capacity versus temperature at zero field for YCo$_3$B$_2$.
\label{Fig-YCoB-Res}}
\end{figure*}

\begin{figure*}[tb]
	\includegraphics[width=0.98\textwidth, trim={4.0cm 67.5cm 2.0cm 8cm},clip]{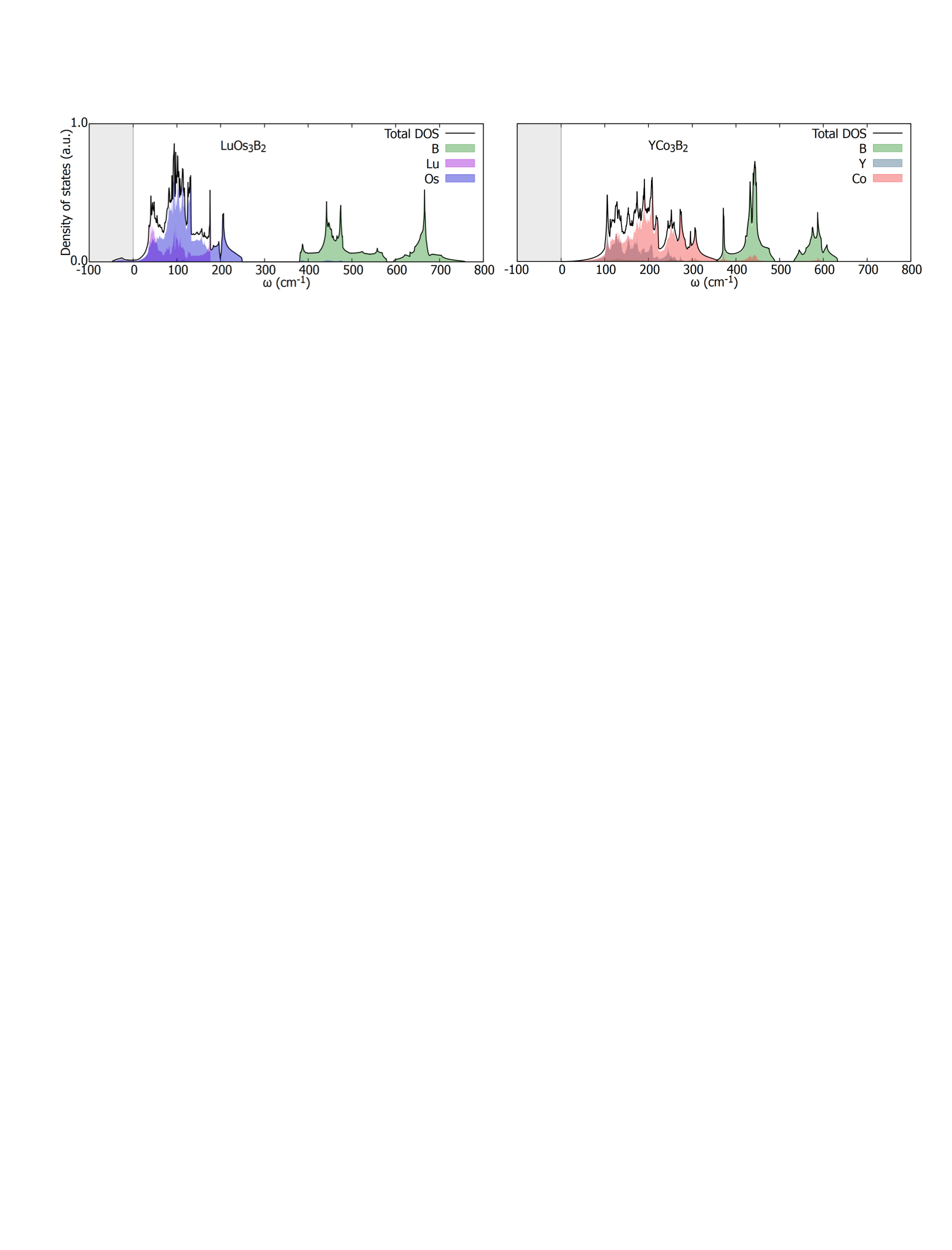}
	\caption{(left panel) Phonon DOS for LuOs$_3$B$_2$. (right panel) Phonon DOS for YCo$_3$B$_2$.}
	\label{fig:phonon-DOS}
\end{figure*}

\begin{figure*}[b]
	\includegraphics[width=0.98\textwidth, trim={4.5cm 53cm 7.5cm 7cm},clip]{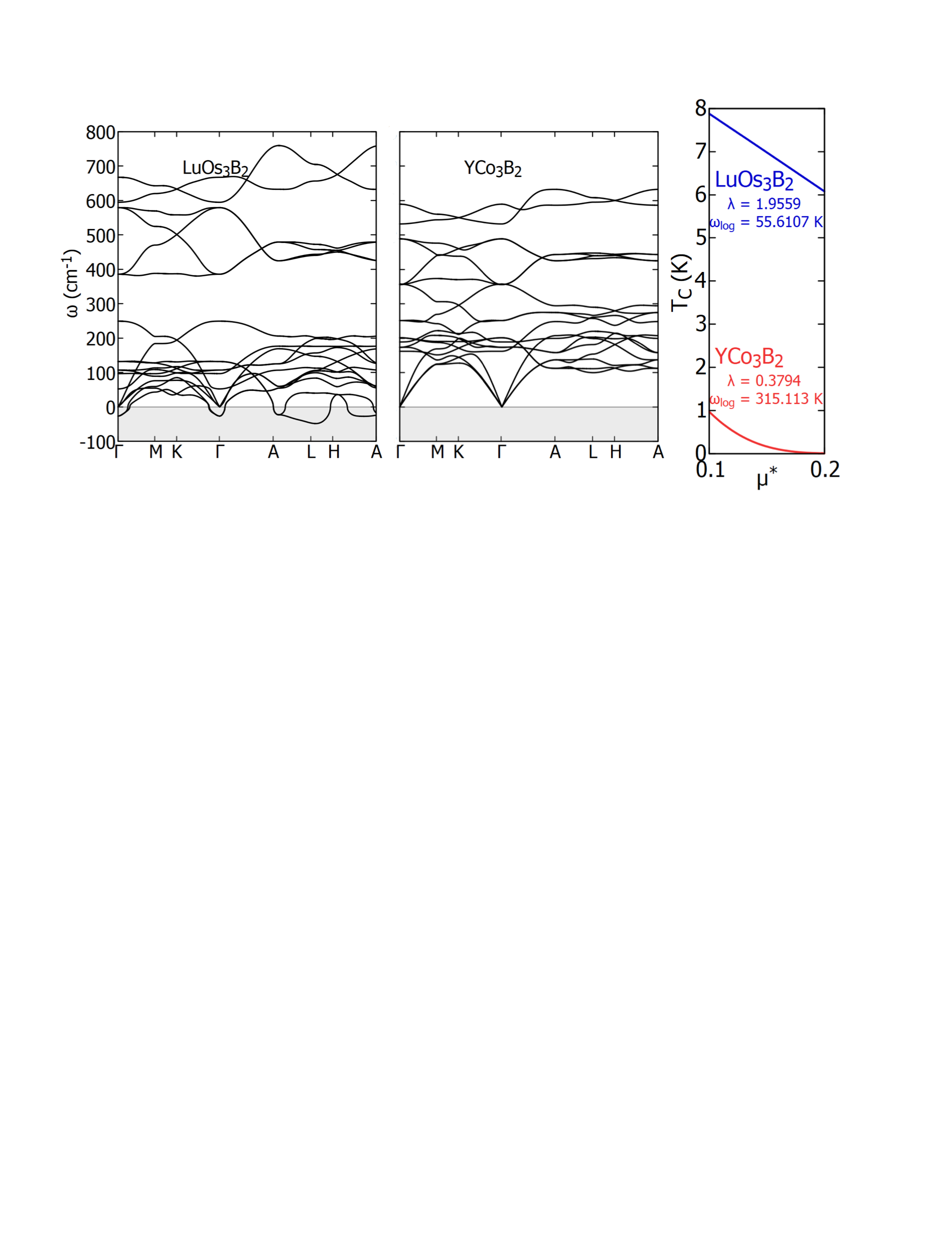}
	\caption{(left panel) Phonon dispersion along high-symmetry lines for LuOs$_3$B$_2$ and (central panel) YCo$_3$B$_2$. (Right panel) Superconducting critical temperature $T_c$(K) as a function of the Coulomb pseudo-potential $\mu^*$.}
	\label{fig:phonon-bands}
\end{figure*}

We discuss the properties of YCo$_3$B$_2$ now.  The resistivity $\rho$ versus temperature $T$ for YCo$_3$B$_2$ measured between $2$ and $300$~K is shown in Fig.~\ref{Fig-YCoB-Res}(a).  The RRR $\approx 10$ indicates a good quality sample.  The $\rho(T)$ shows typical metallic behavior. The $\rho$ data below $\approx 50$~K can be fit by a Fermi-liquid behavior $\rho(T) = \rho_0 + AT^2 + BT^5$.  The fit is shown as the curve through the data in Fig.~\ref{Fig-YCoB-Res}(a) inset. The fit gave the values $\rho_0 = 17.34(7)~\mu\Omega$~cm and $A = 2.2(6) \times 10^{-3}~\mu \Omega$~cm~K$^{-2}$.  The value of $A$, which is related to the strength of the electron-electron scattering, is enhanced and is similar to values found in heavy fermion materials. The heat capacity $C$ for YCo$_3$B$_2$ below $T = 50$~K is shown in Fig.~\ref{Fig-YCoB-Res}(b).  The data below $T \sim 10$~K follows a Fermi-liquid behavior $C/T = \gamma + \beta T^2$. A fit to this expression is shown as the curve through the data in Fig.~\ref{Fig-YCoB-Res}(b) inset and gave the value $\gamma = 12.83(4)$~mJ/mol~K$^2$. The Kadowaki-Woods ratio $A/\gamma^2 \approx 13\mu\Omega$~cm~mol$^2$K$^2$J$^{-2}$, which is much larger than values found for transition metals close to magnetism like Ni, Pd but is close to the value observed for correlated metals likes heavy fermion systems.

\subsection{Phonon Calculations}
The phonon density of states for LuOs$_3$B$_2$ and YCo$_3$B$_2$ are shown in Fig.\ref{fig:phonon-DOS}.  The contribution from different elements is displayed in different colors.   The phonon DOS look quite different for the two materials.  In YCo$_3$B$_2$ there are no imaginary modes, the B states have a gap around $500$~cm$^{-1}$, and part of these B states overlap with the Co states.  In the LuOs$_3$B$_2$ DOS we see imaginary modes.  Also, the B states are well separated from the Os and Lu states.  For both materials the low-frequency phonon modes mostly originate from Y or Lu atoms, while intermediate frequencies are essentially due to Co or Rh atoms; the manifold of low-dispersing bands for both materials is then due to the kagome network. Finally, the B atoms contribute to the high-frequency modes.

The phonon band structure is shown in Fig.~\ref{fig:phonon-bands}.  For LuOs$_3$B$_2$ we see that imaginary modes arise at high symmetry points $\Gamma$, $A$ and $L$.  Imaginary modes in the phonon spectra point to an instability to structural distortions. While the imaginary modes at the Brillouin zone center may be attributed to a numerical approximation artifact, such as the finite size of the super-cell, the mode at the $L$ point drives the system toward a lower-energy configuration and a mildly distorted structure (not shown).
The phonon bands for YCo$_3$B$_2$ are quite flat, similar to what was found for its electronic bands.

From the phonon spectra, we computed the electron-phonon coupling to be $\lambda_{\mathrm{e-ph}} \approx 0.38$ and $1.96$ for YCo$_3$B$_2$ and LuOs$_3$B$_2$ respectively.  The McMillan formula was then used to estimate the superconducting critical temperature $T_c$ \cite{PhysRev.167.331,RevModPhys.62.1027}:
\begin{equation}
    T_c = \frac{\omega_{log}}{1.2} e^{\bigl[\frac{-1.04(1+\lambda)}{\lambda (1-0.62\mu^*) - \mu^*}\bigr]}
\end{equation}
with $\omega_{log}$ being related to the Eliashberg function:
\begin{equation}
    \omega_{log} = e^{\bigl[\frac{2}{\lambda} \int{\frac{d\omega}{\omega} \alpha^2 F(\omega) \log{\omega}} \bigr]}
\end{equation}
While the Coulomb pseudo-potential $\mu^*$ lies in the typical range [0.1 - 0.2], we obtain values for $T_c \approx 6$~K for LuOs$_3$B$_2$ in fair agreement with the experimental results. Our phonon calculations point to intermediate to strong-coupling phonon-mediated superconductivity in LuOs$_3$B$_2$.  On the other hand, the $T_c$ is estimated to be close to $0$ for YCo$_3$B$_2$ consistent with no superconductivity observed down to $2$~K as can be seen from Fig.~\ref{fig:phonon-bands}.\\

\section{Summary and Discussion:}    
The family of kagome metals $RT_3$B$_2$ has a structure built up of kagome planes of $T$ stacked along the $c$-axis with $R$-B planes separating the kagome planes.  We have studied $RT_3$B$_2$ ($R =$ Y, Lu, and $T =$ Co, Os) through measurements of their electronic transport, magnetic susceptibility, and heat capacity and through calculations of their electronic band structure, Fermi surface, and phonon spectrum.  
The electronic structure contains all the features expected for a 2D kagome lattice including a flat band, Dirac bands and van Hove singularities at high symmetry points in the Brillouin zone and the kagome planes made out of Co or Os make a dominant contribution to these features. 
The various measurements point to the importance of electron correlation as seen in the enhanced values of the Wilson ratio ($\sim 3$) and the Kadowaki-Woods ratio ($\sim 48$) for LuOs$_3$B$_2$ and the Kadowaki-Woods ratio ($\sim 13$) for YCo$_3$B$_2$.  

The superconducting properties of LuOs$_3$B$_2$ with a $T_c = 4.8$~K points to Type-II superconductivity with intermediate to strong electron-phonon coupling.  The jump in the heat capacity at $T_c$ is smaller than expectation for a single-gap $s$-wave BCS superconductor, suggesting that LuOs$_3$B$_2$ could be a multi-gap superconductor. Spectroscopic measurements of the gap could be useful to clarify this.

The phonon calculations for LuOs$_3$B$_2$ show imaginary modes, especially at the $L$ high-symmetry point.  Imaginary modes in the phonon spectra indicate an instability to structural distortions and may point to a susceptibility of the system to charge order or CDW like states.  Our measurements did not reveal any such transitions in the temperature range studied.  However, it is possible that LuOs$_3$B$_2$ may be situated at a tipping point and small perturbations like doping or pressure could push the system towards such an instability.
We recall that in LaRh$_3$B$_2$, which we reported on previously showed, no significant features of electronic correlations and no imaginary modes in its phonon spectra.  The fact that we observe both strong correlations and imaginary phonon modes for LuOs$_3$B$_2$ points to an intimate connection between the two. \\
Thus, LuOs$_3$B$_2$ seems to be a good candidate for further study through doping or pressure to try to explore the interplay between correlations and phonon anomalies.

On the other hand in YCo$_3$B$_2$, which has a large density of states just below $E_F$, it may also be interesting to dope the system to lower $E_F$ into the large DOS.  The small electron-phonon coupling calculated for YCo$_3$B$_2$ gives a likely reason for the absence of superconductivity in this material.

\emph{Note Added --} While preparing this draft a report on the synthesis, physical properties, and electronic band structure calculations of LuOs$_3$B$_2$ has appeared \cite{PhysRevB.111.195132}.  Our superconducting and electronic structure properties are in qualitative agreement with those reported in \cite{PhysRevB.111.195132}.  However, we note that this paper reports the synthesis of single phase samples of LuOs$_3$B$_2$ by arc melting whereas we have found LuOs$_3$B$_2$ to be incongruently melting and therefore not possible to obtain in single phase by arc-melting.

\emph{Acknowledgments.--} We thank the X-ray, SEM, and liquid Helium central facilities at IISER Mohali. Y. S. acknowledges support from SERB project CRG/2022/000015 and STARS project STARS-1/240, A. C. acknowledges support from PNRR MUR project PE0000023-NQSTI. A. C., D. D. S. and R. T. acknowledge the Gauss Centre for Supercomputing e.V. (https://www.gauss-centre.eu) for funding this project by providing computing time on the GCS Supercomputer SuperMUC-NG at Leibniz Supercomputing Centre (https://www.lrz.de).

\bibliography{references}

\begin{thebibliography}{51}%
\makeatletter
\providecommand \@ifxundefined [1]{%
 \@ifx{#1\undefined}
}%
\providecommand \@ifnum [1]{%
 \ifnum #1\expandafter \@firstoftwo
 \else \expandafter \@secondoftwo
 \fi
}%
\providecommand \@ifx [1]{%
 \ifx #1\expandafter \@firstoftwo
 \else \expandafter \@secondoftwo
 \fi
}%
\providecommand \natexlab [1]{#1}%
\providecommand \enquote  [1]{``#1''}%
\providecommand \bibnamefont  [1]{#1}%
\providecommand \bibfnamefont [1]{#1}%
\providecommand \citenamefont [1]{#1}%
\providecommand \href@noop [0]{\@secondoftwo}%
\providecommand \href [0]{\begingroup \@sanitize@url \@href}%
\providecommand \@href[1]{\@@startlink{#1}\@@href}%
\providecommand \@@href[1]{\endgroup#1\@@endlink}%
\providecommand \@sanitize@url [0]{\catcode `\\12\catcode `\$12\catcode
  `\&12\catcode `\#12\catcode `\^12\catcode `\_12\catcode `\%12\relax}%
\providecommand \@@startlink[1]{}%
\providecommand \@@endlink[0]{}%
\providecommand \url  [0]{\begingroup\@sanitize@url \@url }%
\providecommand \@url [1]{\endgroup\@href {#1}{\urlprefix }}%
\providecommand \urlprefix  [0]{URL }%
\providecommand \Eprint [0]{\href }%
\providecommand \doibase [0]{https://doi.org/}%
\providecommand \selectlanguage [0]{\@gobble}%
\providecommand \bibinfo  [0]{\@secondoftwo}%
\providecommand \bibfield  [0]{\@secondoftwo}%
\providecommand \translation [1]{[#1]}%
\providecommand \BibitemOpen [0]{}%
\providecommand \bibitemStop [0]{}%
\providecommand \bibitemNoStop [0]{.\EOS\space}%
\providecommand \EOS [0]{\spacefactor3000\relax}%
\providecommand \BibitemShut  [1]{\csname bibitem#1\endcsname}%
\let\auto@bib@innerbib\@empty
\bibitem [{\citenamefont {Balents}(2010)}]{Balents2010}%
  \BibitemOpen
  \bibfield  {author} {\bibinfo {author} {\bibfnamefont {L.}~\bibnamefont
  {Balents}},\ }\bibfield  {title} {\bibinfo {title} {Spin liquids in
  frustrated magnets},\ }\href {https://doi.org/10.1038/nature08917} {\bibfield
   {journal} {\bibinfo  {journal} {Nature}\ }\textbf {\bibinfo {volume}
  {464}},\ \bibinfo {pages} {199} (\bibinfo {year} {2010})}\BibitemShut
  {NoStop}%
\bibitem [{\citenamefont {Savary}\ and\ \citenamefont
  {Balents}(2016)}]{Savary_2017}%
  \BibitemOpen
  \bibfield  {author} {\bibinfo {author} {\bibfnamefont {L.}~\bibnamefont
  {Savary}}\ and\ \bibinfo {author} {\bibfnamefont {L.}~\bibnamefont
  {Balents}},\ }\bibfield  {title} {\bibinfo {title} {Quantum spin liquids: a
  review},\ }\href {https://doi.org/10.1088/0034-4885/80/1/016502} {\bibfield
  {journal} {\bibinfo  {journal} {Reports on Progress in Physics}\ }\textbf
  {\bibinfo {volume} {80}},\ \bibinfo {pages} {016502} (\bibinfo {year}
  {2016})}\BibitemShut {NoStop}%
\bibitem [{\citenamefont {Broholm}\ \emph {et~al.}(2020)\citenamefont
  {Broholm}, \citenamefont {Cava}, \citenamefont {Kivelson}, \citenamefont
  {Nocera}, \citenamefont {Norman},\ and\ \citenamefont
  {Senthil}}]{doi:10.1126/science.aay0668}%
  \BibitemOpen
  \bibfield  {author} {\bibinfo {author} {\bibfnamefont {C.}~\bibnamefont
  {Broholm}}, \bibinfo {author} {\bibfnamefont {R.~J.}\ \bibnamefont {Cava}},
  \bibinfo {author} {\bibfnamefont {S.~A.}\ \bibnamefont {Kivelson}}, \bibinfo
  {author} {\bibfnamefont {D.~G.}\ \bibnamefont {Nocera}}, \bibinfo {author}
  {\bibfnamefont {M.~R.}\ \bibnamefont {Norman}},\ and\ \bibinfo {author}
  {\bibfnamefont {T.}~\bibnamefont {Senthil}},\ }\bibfield  {title} {\bibinfo
  {title} {Quantum spin liquids},\ }\href
  {https://doi.org/10.1126/science.aay0668} {\bibfield  {journal} {\bibinfo
  {journal} {Science}\ }\textbf {\bibinfo {volume} {367}},\ \bibinfo {pages}
  {eaay0668} (\bibinfo {year} {2020})}\BibitemShut {NoStop}%
\bibitem [{\citenamefont {Knolle}\ and\ \citenamefont
  {Moessner}(2019)}]{Knolle:2018xhp}%
  \BibitemOpen
  \bibfield  {author} {\bibinfo {author} {\bibfnamefont {J.}~\bibnamefont
  {Knolle}}\ and\ \bibinfo {author} {\bibfnamefont {R.}~\bibnamefont
  {Moessner}},\ }\bibfield  {title} {\bibinfo {title} {{A Field Guide to Spin
  Liquids}},\ }\href {https://doi.org/10.1146/annurev-conmatphys-031218-013401}
  {\bibfield  {journal} {\bibinfo  {journal} {Ann. Rev. Condensed Matter
  Phys.}\ }\textbf {\bibinfo {volume} {10}},\ \bibinfo {pages} {451} (\bibinfo
  {year} {2019})},\ \Eprint {https://arxiv.org/abs/1804.02037}
  {arXiv:1804.02037 [cond-mat.str-el]} \BibitemShut {NoStop}%
\bibitem [{\citenamefont {Fu}\ \emph {et~al.}(2015)\citenamefont {Fu},
  \citenamefont {Imai}, \citenamefont {Han},\ and\ \citenamefont
  {Lee}}]{doi:10.1126/science.aab2120}%
  \BibitemOpen
  \bibfield  {author} {\bibinfo {author} {\bibfnamefont {M.}~\bibnamefont
  {Fu}}, \bibinfo {author} {\bibfnamefont {T.}~\bibnamefont {Imai}}, \bibinfo
  {author} {\bibfnamefont {T.-H.}\ \bibnamefont {Han}},\ and\ \bibinfo {author}
  {\bibfnamefont {Y.~S.}\ \bibnamefont {Lee}},\ }\bibfield  {title} {\bibinfo
  {title} {Evidence for a gapped spin-liquid ground state in a kagome
  heisenberg antiferromagnet},\ }\href
  {https://doi.org/10.1126/science.aab2120} {\bibfield  {journal} {\bibinfo
  {journal} {Science}\ }\textbf {\bibinfo {volume} {350}},\ \bibinfo {pages}
  {655} (\bibinfo {year} {2015})},\ \Eprint
  {https://arxiv.org/abs/https://www.science.org/doi/pdf/10.1126/science.aab2120}
  {https://www.science.org/doi/pdf/10.1126/science.aab2120} \BibitemShut
  {NoStop}%
\bibitem [{\citenamefont {Han}\ \emph {et~al.}(2012)\citenamefont {Han},
  \citenamefont {Helton}, \citenamefont {Chu}, \citenamefont {Nocera},
  \citenamefont {Rodriguez-Rivera}, \citenamefont {Broholm},\ and\
  \citenamefont {Lee}}]{Han2012}%
  \BibitemOpen
  \bibfield  {author} {\bibinfo {author} {\bibfnamefont {T.-H.}\ \bibnamefont
  {Han}}, \bibinfo {author} {\bibfnamefont {J.~S.}\ \bibnamefont {Helton}},
  \bibinfo {author} {\bibfnamefont {S.}~\bibnamefont {Chu}}, \bibinfo {author}
  {\bibfnamefont {D.~G.}\ \bibnamefont {Nocera}}, \bibinfo {author}
  {\bibfnamefont {J.~A.}\ \bibnamefont {Rodriguez-Rivera}}, \bibinfo {author}
  {\bibfnamefont {C.}~\bibnamefont {Broholm}},\ and\ \bibinfo {author}
  {\bibfnamefont {Y.~S.}\ \bibnamefont {Lee}},\ }\bibfield  {title} {\bibinfo
  {title} {Fractionalized excitations in the spin-liquid state of a
  kagome-lattice antiferromagnet},\ }\href
  {https://doi.org/10.1038/nature11659} {\bibfield  {journal} {\bibinfo
  {journal} {Nature}\ }\textbf {\bibinfo {volume} {492}},\ \bibinfo {pages}
  {406} (\bibinfo {year} {2012})}\BibitemShut {NoStop}%
\bibitem [{\citenamefont {Balz}\ \emph {et~al.}(2016)\citenamefont {Balz},
  \citenamefont {Lake}, \citenamefont {Reuther}, \citenamefont {Luetkens},
  \citenamefont {Sch{\"o}nemann}, \citenamefont {Herrmannsd{\"o}rfer},
  \citenamefont {Singh}, \citenamefont {Nazmul~Islam}, \citenamefont {Wheeler},
  \citenamefont {Rodriguez-Rivera}, \citenamefont {Guidi}, \citenamefont
  {Simeoni}, \citenamefont {Baines},\ and\ \citenamefont {Ryll}}]{Balz2016}%
  \BibitemOpen
  \bibfield  {author} {\bibinfo {author} {\bibfnamefont {C.}~\bibnamefont
  {Balz}}, \bibinfo {author} {\bibfnamefont {B.}~\bibnamefont {Lake}}, \bibinfo
  {author} {\bibfnamefont {J.}~\bibnamefont {Reuther}}, \bibinfo {author}
  {\bibfnamefont {H.}~\bibnamefont {Luetkens}}, \bibinfo {author}
  {\bibfnamefont {R.}~\bibnamefont {Sch{\"o}nemann}}, \bibinfo {author}
  {\bibfnamefont {T.}~\bibnamefont {Herrmannsd{\"o}rfer}}, \bibinfo {author}
  {\bibfnamefont {Y.}~\bibnamefont {Singh}}, \bibinfo {author} {\bibfnamefont
  {A.~T.~M.}\ \bibnamefont {Nazmul~Islam}}, \bibinfo {author} {\bibfnamefont
  {E.~M.}\ \bibnamefont {Wheeler}}, \bibinfo {author} {\bibfnamefont
  {J.}~\bibnamefont {Rodriguez-Rivera}}, \bibinfo {author} {\bibfnamefont
  {T.}~\bibnamefont {Guidi}}, \bibinfo {author} {\bibfnamefont
  {G.}~\bibnamefont {Simeoni}}, \bibinfo {author} {\bibfnamefont
  {C.}~\bibnamefont {Baines}},\ and\ \bibinfo {author} {\bibfnamefont
  {H.}~\bibnamefont {Ryll}},\ }\bibfield  {title} {\bibinfo {title} {Physical
  realization of a quantum spin liquid based on a complex frustration
  mechanism},\ }\href {https://doi.org/10.1038/nphys3826} {\bibfield  {journal}
  {\bibinfo  {journal} {Nature Physics}\ }\textbf {\bibinfo {volume} {12}},\
  \bibinfo {pages} {942} (\bibinfo {year} {2016})}\BibitemShut {NoStop}%
\bibitem [{\citenamefont {Okamoto}\ \emph {et~al.}(2007)\citenamefont
  {Okamoto}, \citenamefont {Nohara}, \citenamefont {Aruga-Katori},\ and\
  \citenamefont {Takagi}}]{PhysRevLett.99.137207}%
  \BibitemOpen
  \bibfield  {author} {\bibinfo {author} {\bibfnamefont {Y.}~\bibnamefont
  {Okamoto}}, \bibinfo {author} {\bibfnamefont {M.}~\bibnamefont {Nohara}},
  \bibinfo {author} {\bibfnamefont {H.}~\bibnamefont {Aruga-Katori}},\ and\
  \bibinfo {author} {\bibfnamefont {H.}~\bibnamefont {Takagi}},\ }\bibfield
  {title} {\bibinfo {title} {Spin-liquid state in the $s=1/2$ hyperkagome
  antiferromagnet ${\mathrm{na}}_{4}{\mathrm{ir}}_{3}{\mathrm{o}}_{8}$},\
  }\href {https://doi.org/10.1103/PhysRevLett.99.137207} {\bibfield  {journal}
  {\bibinfo  {journal} {Phys. Rev. Lett.}\ }\textbf {\bibinfo {volume} {99}},\
  \bibinfo {pages} {137207} (\bibinfo {year} {2007})}\BibitemShut {NoStop}%
\bibitem [{\citenamefont {Singh}\ \emph {et~al.}(2013)\citenamefont {Singh},
  \citenamefont {Tokiwa}, \citenamefont {Dong},\ and\ \citenamefont
  {Gegenwart}}]{PhysRevB.88.220413}%
  \BibitemOpen
  \bibfield  {author} {\bibinfo {author} {\bibfnamefont {Y.}~\bibnamefont
  {Singh}}, \bibinfo {author} {\bibfnamefont {Y.}~\bibnamefont {Tokiwa}},
  \bibinfo {author} {\bibfnamefont {J.}~\bibnamefont {Dong}},\ and\ \bibinfo
  {author} {\bibfnamefont {P.}~\bibnamefont {Gegenwart}},\ }\bibfield  {title}
  {\bibinfo {title} {Spin liquid close to a quantum critical point in
  na${}_{4}$ir${}_{3}$o${}_{8}$},\ }\href
  {https://doi.org/10.1103/PhysRevB.88.220413} {\bibfield  {journal} {\bibinfo
  {journal} {Phys. Rev. B}\ }\textbf {\bibinfo {volume} {88}},\ \bibinfo
  {pages} {220413} (\bibinfo {year} {2013})}\BibitemShut {NoStop}%
\bibitem [{\citenamefont {Mazin}\ \emph {et~al.}(2014)\citenamefont {Mazin},
  \citenamefont {Jeschke}, \citenamefont {Lechermann}, \citenamefont {Lee},
  \citenamefont {Fink}, \citenamefont {Thomale},\ and\ \citenamefont
  {Valent{\'i}}}]{9e66d15ca73e41e8a8a2746de65fd232}%
  \BibitemOpen
  \bibfield  {author} {\bibinfo {author} {\bibfnamefont {I.}~\bibnamefont
  {Mazin}}, \bibinfo {author} {\bibfnamefont {H.}~\bibnamefont {Jeschke}},
  \bibinfo {author} {\bibfnamefont {F.}~\bibnamefont {Lechermann}}, \bibinfo
  {author} {\bibfnamefont {H.}~\bibnamefont {Lee}}, \bibinfo {author}
  {\bibfnamefont {M.}~\bibnamefont {Fink}}, \bibinfo {author} {\bibfnamefont
  {R.}~\bibnamefont {Thomale}},\ and\ \bibinfo {author} {\bibfnamefont
  {R.}~\bibnamefont {Valent{\'i}}},\ }\bibfield  {title} {{\selectlanguage
  {English}\bibinfo {title} {Theoretical prediction of a strongly correlated
  dirac metal}},\ }\bibfield  {journal} {\bibinfo  {journal} {Nature
  communications}\ }\textbf {\bibinfo {volume} {5}},\ \href
  {https://doi.org/10.1038/ncomms5261} {10.1038/ncomms5261} (\bibinfo {year}
  {2014}),\ \bibinfo {note} {funding Information: We acknowledge useful
  discussions with C. Krellner, C. Platt, G. Khalliulin, A. Chubukov and C.
  Piefke, C.S. Hellberg, M. Taillefumier and R. Moessner. I.I.M. is supported
  by ONR through the NRL basic research program and in part by the A. von
  Humboldt Foundation. H.O.J., F.L. and R.V. are supported by DFG-FOR1346 and
  DFG-SFB/TR49 (H.O.J., R.V.). R.T. is supported by the European Research
  Council through the grant TOPOLECTRICS, ERC-StG-336012.}\BibitemShut {Stop}%
\bibitem [{\citenamefont {Kang}\ \emph
  {et~al.}(2020{\natexlab{a}})\citenamefont {Kang}, \citenamefont {Ye},
  \citenamefont {Fang}, \citenamefont {You}, \citenamefont {Levitan},
  \citenamefont {Han}, \citenamefont {Facio}, \citenamefont {Jozwiak},
  \citenamefont {Bostwick}, \citenamefont {Rotenberg}, \citenamefont {Chan},
  \citenamefont {McDonald}, \citenamefont {Graf}, \citenamefont {Kaznatcheev},
  \citenamefont {Vescovo}, \citenamefont {Bell}, \citenamefont {Kaxiras},
  \citenamefont {{van den Brink}}, \citenamefont {Richter}, \citenamefont
  {{Prasad Ghimire}}, \citenamefont {Checkelsky},\ and\ \citenamefont
  {Comin}}]{95810bfd2245410db75382c1581ffe9e}%
  \BibitemOpen
  \bibfield  {author} {\bibinfo {author} {\bibfnamefont {M.}~\bibnamefont
  {Kang}}, \bibinfo {author} {\bibfnamefont {L.}~\bibnamefont {Ye}}, \bibinfo
  {author} {\bibfnamefont {S.}~\bibnamefont {Fang}}, \bibinfo {author}
  {\bibfnamefont {J.}~\bibnamefont {You}}, \bibinfo {author} {\bibfnamefont
  {A.}~\bibnamefont {Levitan}}, \bibinfo {author} {\bibfnamefont
  {M.}~\bibnamefont {Han}}, \bibinfo {author} {\bibfnamefont {J.}~\bibnamefont
  {Facio}}, \bibinfo {author} {\bibfnamefont {C.}~\bibnamefont {Jozwiak}},
  \bibinfo {author} {\bibfnamefont {A.}~\bibnamefont {Bostwick}}, \bibinfo
  {author} {\bibfnamefont {E.}~\bibnamefont {Rotenberg}}, \bibinfo {author}
  {\bibfnamefont {M.}~\bibnamefont {Chan}}, \bibinfo {author} {\bibfnamefont
  {R.}~\bibnamefont {McDonald}}, \bibinfo {author} {\bibfnamefont
  {D.}~\bibnamefont {Graf}}, \bibinfo {author} {\bibfnamefont {K.}~\bibnamefont
  {Kaznatcheev}}, \bibinfo {author} {\bibfnamefont {E.}~\bibnamefont
  {Vescovo}}, \bibinfo {author} {\bibfnamefont {D.}~\bibnamefont {Bell}},
  \bibinfo {author} {\bibfnamefont {E.}~\bibnamefont {Kaxiras}}, \bibinfo
  {author} {\bibfnamefont {J.}~\bibnamefont {{van den Brink}}}, \bibinfo
  {author} {\bibfnamefont {M.}~\bibnamefont {Richter}}, \bibinfo {author}
  {\bibfnamefont {M.}~\bibnamefont {{Prasad Ghimire}}}, \bibinfo {author}
  {\bibfnamefont {J.}~\bibnamefont {Checkelsky}},\ and\ \bibinfo {author}
  {\bibfnamefont {R.}~\bibnamefont {Comin}},\ }\bibfield  {title}
  {{\selectlanguage {English}\bibinfo {title} {Dirac fermions and flat bands in
  the ideal kagome metal fesn}},\ }\href
  {https://doi.org/10.1038/s41563-019-0531-0} {\bibfield  {journal} {\bibinfo
  {journal} {Nature Materials}\ }\textbf {\bibinfo {volume} {19}},\ \bibinfo
  {pages} {163} (\bibinfo {year} {2020}{\natexlab{a}})},\ \bibinfo {note}
  {publisher Copyright: {\textcopyright} 2019, The Author(s), under exclusive
  licence to Springer Nature Limited.}\BibitemShut {Stop}%
\bibitem [{\citenamefont {Kang}\ \emph
  {et~al.}(2020{\natexlab{b}})\citenamefont {Kang}, \citenamefont {Fang},
  \citenamefont {Ye}, \citenamefont {Po}, \citenamefont {Denlinger},
  \citenamefont {Jozwiak}, \citenamefont {Bostwick}, \citenamefont {Rotenberg},
  \citenamefont {Kaxiras}, \citenamefont {Checkelsky},\ and\ \citenamefont
  {Comin}}]{0e7cb127b339477390e474f0da9235e5}%
  \BibitemOpen
  \bibfield  {author} {\bibinfo {author} {\bibfnamefont {M.}~\bibnamefont
  {Kang}}, \bibinfo {author} {\bibfnamefont {S.}~\bibnamefont {Fang}}, \bibinfo
  {author} {\bibfnamefont {L.}~\bibnamefont {Ye}}, \bibinfo {author}
  {\bibfnamefont {H.}~\bibnamefont {Po}}, \bibinfo {author} {\bibfnamefont
  {J.}~\bibnamefont {Denlinger}}, \bibinfo {author} {\bibfnamefont
  {C.}~\bibnamefont {Jozwiak}}, \bibinfo {author} {\bibfnamefont
  {A.}~\bibnamefont {Bostwick}}, \bibinfo {author} {\bibfnamefont
  {E.}~\bibnamefont {Rotenberg}}, \bibinfo {author} {\bibfnamefont
  {E.}~\bibnamefont {Kaxiras}}, \bibinfo {author} {\bibfnamefont
  {J.}~\bibnamefont {Checkelsky}},\ and\ \bibinfo {author} {\bibfnamefont
  {R.}~\bibnamefont {Comin}},\ }\bibfield  {title} {{\selectlanguage
  {English}\bibinfo {title} {Topological flat bands in frustrated kagome
  lattice cosn}},\ }\bibfield  {journal} {\bibinfo  {journal} {Nature
  Communications}\ }\textbf {\bibinfo {volume} {11}},\ \href
  {https://doi.org/10.1038/s41467-020-17465-1} {10.1038/s41467-020-17465-1}
  (\bibinfo {year} {2020}{\natexlab{b}}),\ \bibinfo {note} {publisher
  Copyright: {\textcopyright} 2020, The Author(s).}\BibitemShut {Stop}%
\bibitem [{\citenamefont {Li}\ \emph {et~al.}(2021)\citenamefont {Li},
  \citenamefont {Wang}, \citenamefont {Wang}, \citenamefont {Yuan},
  \citenamefont {Song}, \citenamefont {Lou}, \citenamefont {Liu}, \citenamefont
  {Huang}, \citenamefont {Liu}, \citenamefont {Lei}, \citenamefont {Yin},\ and\
  \citenamefont {Wang}}]{Li2021}%
  \BibitemOpen
  \bibfield  {author} {\bibinfo {author} {\bibfnamefont {M.}~\bibnamefont
  {Li}}, \bibinfo {author} {\bibfnamefont {Q.}~\bibnamefont {Wang}}, \bibinfo
  {author} {\bibfnamefont {G.}~\bibnamefont {Wang}}, \bibinfo {author}
  {\bibfnamefont {Z.}~\bibnamefont {Yuan}}, \bibinfo {author} {\bibfnamefont
  {W.}~\bibnamefont {Song}}, \bibinfo {author} {\bibfnamefont {R.}~\bibnamefont
  {Lou}}, \bibinfo {author} {\bibfnamefont {Z.}~\bibnamefont {Liu}}, \bibinfo
  {author} {\bibfnamefont {Y.}~\bibnamefont {Huang}}, \bibinfo {author}
  {\bibfnamefont {Z.}~\bibnamefont {Liu}}, \bibinfo {author} {\bibfnamefont
  {H.}~\bibnamefont {Lei}}, \bibinfo {author} {\bibfnamefont {Z.}~\bibnamefont
  {Yin}},\ and\ \bibinfo {author} {\bibfnamefont {S.}~\bibnamefont {Wang}},\
  }\bibfield  {title} {\bibinfo {title} {Dirac cone, flat band and saddle point
  in kagome magnet ymn6sn6},\ }\href
  {https://doi.org/10.1038/s41467-021-23536-8} {\bibfield  {journal} {\bibinfo
  {journal} {Nature Communications}\ }\textbf {\bibinfo {volume} {12}},\
  \bibinfo {pages} {3129} (\bibinfo {year} {2021})}\BibitemShut {NoStop}%
\bibitem [{\citenamefont {Ortiz}\ \emph {et~al.}(2019)\citenamefont {Ortiz},
  \citenamefont {Gomes}, \citenamefont {Morey}, \citenamefont {Winiarski},
  \citenamefont {Bordelon}, \citenamefont {Mangum}, \citenamefont {Oswald},
  \citenamefont {Rodriguez-Rivera}, \citenamefont {Neilson}, \citenamefont
  {Wilson}, \citenamefont {Ertekin}, \citenamefont {McQueen},\ and\
  \citenamefont {Toberer}}]{PhysRevMaterials.3.094407}%
  \BibitemOpen
  \bibfield  {author} {\bibinfo {author} {\bibfnamefont {B.~R.}\ \bibnamefont
  {Ortiz}}, \bibinfo {author} {\bibfnamefont {L.~C.}\ \bibnamefont {Gomes}},
  \bibinfo {author} {\bibfnamefont {J.~R.}\ \bibnamefont {Morey}}, \bibinfo
  {author} {\bibfnamefont {M.}~\bibnamefont {Winiarski}}, \bibinfo {author}
  {\bibfnamefont {M.}~\bibnamefont {Bordelon}}, \bibinfo {author}
  {\bibfnamefont {J.~S.}\ \bibnamefont {Mangum}}, \bibinfo {author}
  {\bibfnamefont {I.~W.~H.}\ \bibnamefont {Oswald}}, \bibinfo {author}
  {\bibfnamefont {J.~A.}\ \bibnamefont {Rodriguez-Rivera}}, \bibinfo {author}
  {\bibfnamefont {J.~R.}\ \bibnamefont {Neilson}}, \bibinfo {author}
  {\bibfnamefont {S.~D.}\ \bibnamefont {Wilson}}, \bibinfo {author}
  {\bibfnamefont {E.}~\bibnamefont {Ertekin}}, \bibinfo {author} {\bibfnamefont
  {T.~M.}\ \bibnamefont {McQueen}},\ and\ \bibinfo {author} {\bibfnamefont
  {E.~S.}\ \bibnamefont {Toberer}},\ }\bibfield  {title} {\bibinfo {title} {New
  kagome prototype materials: discovery of
  ${\mathrm{kv}}_{3}{\mathrm{sb}}_{5},{\mathrm{rbv}}_{3}{\mathrm{sb}}_{5}$, and
  ${\mathrm{csv}}_{3}{\mathrm{sb}}_{5}$},\ }\href
  {https://doi.org/10.1103/PhysRevMaterials.3.094407} {\bibfield  {journal}
  {\bibinfo  {journal} {Phys. Rev. Mater.}\ }\textbf {\bibinfo {volume} {3}},\
  \bibinfo {pages} {094407} (\bibinfo {year} {2019})}\BibitemShut {NoStop}%
\bibitem [{\citenamefont {Ortiz}\ \emph {et~al.}(2020)\citenamefont {Ortiz},
  \citenamefont {Teicher}, \citenamefont {Hu}, \citenamefont {Zuo},
  \citenamefont {Sarte}, \citenamefont {Schueller}, \citenamefont {Abeykoon},
  \citenamefont {Krogstad}, \citenamefont {Rosenkranz}, \citenamefont {Osborn},
  \citenamefont {Seshadri}, \citenamefont {Balents}, \citenamefont {He},\ and\
  \citenamefont {Wilson}}]{PhysRevLett.125.247002}%
  \BibitemOpen
  \bibfield  {author} {\bibinfo {author} {\bibfnamefont {B.~R.}\ \bibnamefont
  {Ortiz}}, \bibinfo {author} {\bibfnamefont {S.~M.~L.}\ \bibnamefont
  {Teicher}}, \bibinfo {author} {\bibfnamefont {Y.}~\bibnamefont {Hu}},
  \bibinfo {author} {\bibfnamefont {J.~L.}\ \bibnamefont {Zuo}}, \bibinfo
  {author} {\bibfnamefont {P.~M.}\ \bibnamefont {Sarte}}, \bibinfo {author}
  {\bibfnamefont {E.~C.}\ \bibnamefont {Schueller}}, \bibinfo {author}
  {\bibfnamefont {A.~M.~M.}\ \bibnamefont {Abeykoon}}, \bibinfo {author}
  {\bibfnamefont {M.~J.}\ \bibnamefont {Krogstad}}, \bibinfo {author}
  {\bibfnamefont {S.}~\bibnamefont {Rosenkranz}}, \bibinfo {author}
  {\bibfnamefont {R.}~\bibnamefont {Osborn}}, \bibinfo {author} {\bibfnamefont
  {R.}~\bibnamefont {Seshadri}}, \bibinfo {author} {\bibfnamefont
  {L.}~\bibnamefont {Balents}}, \bibinfo {author} {\bibfnamefont
  {J.}~\bibnamefont {He}},\ and\ \bibinfo {author} {\bibfnamefont {S.~D.}\
  \bibnamefont {Wilson}},\ }\bibfield  {title} {\bibinfo {title}
  {$\mathrm{Cs}{\mathrm{v}}_{3}{\mathrm{sb}}_{5}$: A ${\mathbb{z}}_{2}$
  topological kagome metal with a superconducting ground state},\ }\href
  {https://doi.org/10.1103/PhysRevLett.125.247002} {\bibfield  {journal}
  {\bibinfo  {journal} {Phys. Rev. Lett.}\ }\textbf {\bibinfo {volume} {125}},\
  \bibinfo {pages} {247002} (\bibinfo {year} {2020})}\BibitemShut {NoStop}%
\bibitem [{\citenamefont {Ortiz}\ \emph {et~al.}(2021)\citenamefont {Ortiz},
  \citenamefont {Sarte}, \citenamefont {Kenney}, \citenamefont {Graf},
  \citenamefont {Teicher}, \citenamefont {Seshadri},\ and\ \citenamefont
  {Wilson}}]{PhysRevMaterials.5.034801}%
  \BibitemOpen
  \bibfield  {author} {\bibinfo {author} {\bibfnamefont {B.~R.}\ \bibnamefont
  {Ortiz}}, \bibinfo {author} {\bibfnamefont {P.~M.}\ \bibnamefont {Sarte}},
  \bibinfo {author} {\bibfnamefont {E.~M.}\ \bibnamefont {Kenney}}, \bibinfo
  {author} {\bibfnamefont {M.~J.}\ \bibnamefont {Graf}}, \bibinfo {author}
  {\bibfnamefont {S.~M.~L.}\ \bibnamefont {Teicher}}, \bibinfo {author}
  {\bibfnamefont {R.}~\bibnamefont {Seshadri}},\ and\ \bibinfo {author}
  {\bibfnamefont {S.~D.}\ \bibnamefont {Wilson}},\ }\bibfield  {title}
  {\bibinfo {title} {Superconductivity in the ${\mathbb{z}}_{2}$ kagome metal
  ${\mathrm{kv}}_{3}{\mathrm{sb}}_{5}$},\ }\href
  {https://doi.org/10.1103/PhysRevMaterials.5.034801} {\bibfield  {journal}
  {\bibinfo  {journal} {Phys. Rev. Mater.}\ }\textbf {\bibinfo {volume} {5}},\
  \bibinfo {pages} {034801} (\bibinfo {year} {2021})}\BibitemShut {NoStop}%
\bibitem [{\citenamefont {Yang}\ \emph {et~al.}(2020)\citenamefont {Yang},
  \citenamefont {Wang}, \citenamefont {Ortiz}, \citenamefont {Liu},
  \citenamefont {Gayles}, \citenamefont {Derunova}, \citenamefont
  {Gonzalez-Hernandez}, \citenamefont {Šmejkal}, \citenamefont {Chen},
  \citenamefont {Parkin}, \citenamefont {Wilson}, \citenamefont {Toberer},
  \citenamefont {McQueen},\ and\ \citenamefont
  {Ali}}]{doi:10.1126/sciadv.abb6003}%
  \BibitemOpen
  \bibfield  {author} {\bibinfo {author} {\bibfnamefont {S.-Y.}\ \bibnamefont
  {Yang}}, \bibinfo {author} {\bibfnamefont {Y.}~\bibnamefont {Wang}}, \bibinfo
  {author} {\bibfnamefont {B.~R.}\ \bibnamefont {Ortiz}}, \bibinfo {author}
  {\bibfnamefont {D.}~\bibnamefont {Liu}}, \bibinfo {author} {\bibfnamefont
  {J.}~\bibnamefont {Gayles}}, \bibinfo {author} {\bibfnamefont
  {E.}~\bibnamefont {Derunova}}, \bibinfo {author} {\bibfnamefont
  {R.}~\bibnamefont {Gonzalez-Hernandez}}, \bibinfo {author} {\bibfnamefont
  {L.}~\bibnamefont {Šmejkal}}, \bibinfo {author} {\bibfnamefont
  {Y.}~\bibnamefont {Chen}}, \bibinfo {author} {\bibfnamefont {S.~S.~P.}\
  \bibnamefont {Parkin}}, \bibinfo {author} {\bibfnamefont {S.~D.}\
  \bibnamefont {Wilson}}, \bibinfo {author} {\bibfnamefont {E.~S.}\
  \bibnamefont {Toberer}}, \bibinfo {author} {\bibfnamefont {T.}~\bibnamefont
  {McQueen}},\ and\ \bibinfo {author} {\bibfnamefont {M.~N.}\ \bibnamefont
  {Ali}},\ }\bibfield  {title} {\bibinfo {title} {Giant, unconventional
  anomalous hall effect in the metallic frustrated magnet candidate,
  kv<sub>3</sub>sb<sub>5</sub>},\ }\href
  {https://doi.org/10.1126/sciadv.abb6003} {\bibfield  {journal} {\bibinfo
  {journal} {Science Advances}\ }\textbf {\bibinfo {volume} {6}},\ \bibinfo
  {pages} {eabb6003} (\bibinfo {year} {2020})}\BibitemShut {NoStop}%
\bibitem [{\citenamefont {Ku}\ \emph {et~al.}(1980)\citenamefont {Ku},
  \citenamefont {Meisner}, \citenamefont {Acker},\ and\ \citenamefont
  {Johnston}}]{KU198091}%
  \BibitemOpen
  \bibfield  {author} {\bibinfo {author} {\bibfnamefont {H.}~\bibnamefont
  {Ku}}, \bibinfo {author} {\bibfnamefont {G.}~\bibnamefont {Meisner}},
  \bibinfo {author} {\bibfnamefont {F.}~\bibnamefont {Acker}},\ and\ \bibinfo
  {author} {\bibfnamefont {D.}~\bibnamefont {Johnston}},\ }\bibfield  {title}
  {\bibinfo {title} {Superconducting and magnetic properties of new ternary
  borides with the ceco3b2-type structure},\ }\href
  {https://doi.org/https://doi.org/10.1016/0038-1098(80)90221-5} {\bibfield
  {journal} {\bibinfo  {journal} {Solid State Communications}\ }\textbf
  {\bibinfo {volume} {35}},\ \bibinfo {pages} {91} (\bibinfo {year}
  {1980})}\BibitemShut {NoStop}%
\bibitem [{\citenamefont {Barz}(1980)}]{BARZ19801489}%
  \BibitemOpen
  \bibfield  {author} {\bibinfo {author} {\bibfnamefont {H.}~\bibnamefont
  {Barz}},\ }\bibfield  {title} {\bibinfo {title} {New ternary superconductors
  with silicon},\ }\href
  {https://doi.org/https://doi.org/10.1016/0025-5408(80)90107-5} {\bibfield
  {journal} {\bibinfo  {journal} {Materials Research Bulletin}\ }\textbf
  {\bibinfo {volume} {15}},\ \bibinfo {pages} {1489} (\bibinfo {year}
  {1980})}\BibitemShut {NoStop}%
\bibitem [{\citenamefont {Vandenberg}\ and\ \citenamefont
  {Barz}(1980)}]{VANDENBERG19801493}%
  \BibitemOpen
  \bibfield  {author} {\bibinfo {author} {\bibfnamefont {J.}~\bibnamefont
  {Vandenberg}}\ and\ \bibinfo {author} {\bibfnamefont {H.}~\bibnamefont
  {Barz}},\ }\bibfield  {title} {\bibinfo {title} {The crystal structure of a
  new ternary silicide in the system rare-earth-ruthenium-silicon},\ }\href
  {https://doi.org/https://doi.org/10.1016/0025-5408(80)90108-7} {\bibfield
  {journal} {\bibinfo  {journal} {Materials Research Bulletin}\ }\textbf
  {\bibinfo {volume} {15}},\ \bibinfo {pages} {1493} (\bibinfo {year}
  {1980})}\BibitemShut {NoStop}%
\bibitem [{\citenamefont {Malik}\ \emph {et~al.}(1985)\citenamefont {Malik},
  \citenamefont {Umarji}, \citenamefont {Shenoy}, \citenamefont {Aldred},\ and\
  \citenamefont {Niarchos}}]{PhysRevB.32.4742}%
  \BibitemOpen
  \bibfield  {author} {\bibinfo {author} {\bibfnamefont {S.~K.}\ \bibnamefont
  {Malik}}, \bibinfo {author} {\bibfnamefont {A.~M.}\ \bibnamefont {Umarji}},
  \bibinfo {author} {\bibfnamefont {G.~K.}\ \bibnamefont {Shenoy}}, \bibinfo
  {author} {\bibfnamefont {A.~T.}\ \bibnamefont {Aldred}},\ and\ \bibinfo
  {author} {\bibfnamefont {D.~G.}\ \bibnamefont {Niarchos}},\ }\bibfield
  {title} {\bibinfo {title} {Magnetism and superconductivity in the system
  ${\mathrm{ce}}_{1\mathrm{\ensuremath{-}}\mathrm{x}}$${\mathrm{la}}_{\mathrm{x}}$${\mathrm{rh}}_{3}$${\mathrm{b}}_{2}$},\
  }\href {https://doi.org/10.1103/PhysRevB.32.4742} {\bibfield  {journal}
  {\bibinfo  {journal} {Phys. Rev. B}\ }\textbf {\bibinfo {volume} {32}},\
  \bibinfo {pages} {4742} (\bibinfo {year} {1985})}\BibitemShut {NoStop}%
\bibitem [{\citenamefont {Athreya}\ \emph {et~al.}(1985)\citenamefont
  {Athreya}, \citenamefont {Hausermann-Berg}, \citenamefont {Shelton},
  \citenamefont {Malik}, \citenamefont {Umarji},\ and\ \citenamefont
  {Shenoy}}]{ATHREYA1985330}%
  \BibitemOpen
  \bibfield  {author} {\bibinfo {author} {\bibfnamefont {K.}~\bibnamefont
  {Athreya}}, \bibinfo {author} {\bibfnamefont {L.}~\bibnamefont
  {Hausermann-Berg}}, \bibinfo {author} {\bibfnamefont {R.}~\bibnamefont
  {Shelton}}, \bibinfo {author} {\bibfnamefont {S.}~\bibnamefont {Malik}},
  \bibinfo {author} {\bibfnamefont {A.}~\bibnamefont {Umarji}},\ and\ \bibinfo
  {author} {\bibfnamefont {G.}~\bibnamefont {Shenoy}},\ }\bibfield  {title}
  {\bibinfo {title} {Superconductivity in the ternary borides ceos3b2 and
  ceru3b2: Magnetic susceptibility and specific heat measurements},\ }\href
  {https://doi.org/https://doi.org/10.1016/0375-9601(85)90177-X} {\bibfield
  {journal} {\bibinfo  {journal} {Physics Letters A}\ }\textbf {\bibinfo
  {volume} {113}},\ \bibinfo {pages} {330} (\bibinfo {year}
  {1985})}\BibitemShut {NoStop}%
\bibitem [{\citenamefont {Rauchschwalbe}\ \emph {et~al.}(1984)\citenamefont
  {Rauchschwalbe}, \citenamefont {Lieke}, \citenamefont {Steglich},
  \citenamefont {Godart}, \citenamefont {Gupta},\ and\ \citenamefont
  {Parks}}]{PhysRevB.30.444}%
  \BibitemOpen
  \bibfield  {author} {\bibinfo {author} {\bibfnamefont {U.}~\bibnamefont
  {Rauchschwalbe}}, \bibinfo {author} {\bibfnamefont {W.}~\bibnamefont
  {Lieke}}, \bibinfo {author} {\bibfnamefont {F.}~\bibnamefont {Steglich}},
  \bibinfo {author} {\bibfnamefont {C.}~\bibnamefont {Godart}}, \bibinfo
  {author} {\bibfnamefont {L.~C.}\ \bibnamefont {Gupta}},\ and\ \bibinfo
  {author} {\bibfnamefont {R.~D.}\ \bibnamefont {Parks}},\ }\bibfield  {title}
  {\bibinfo {title} {Superconductivity in a mixed-valent system:
  ${\mathrm{ceru}}_{3}$${\mathrm{si}}_{2}$},\ }\href
  {https://doi.org/10.1103/PhysRevB.30.444} {\bibfield  {journal} {\bibinfo
  {journal} {Phys. Rev. B}\ }\textbf {\bibinfo {volume} {30}},\ \bibinfo
  {pages} {444} (\bibinfo {year} {1984})}\BibitemShut {NoStop}%
\bibitem [{\citenamefont {Li}\ \emph {et~al.}(2011)\citenamefont {Li},
  \citenamefont {Zeng}, \citenamefont {Wan}, \citenamefont {Tao}, \citenamefont
  {Han}, \citenamefont {Yang}, \citenamefont {Wang},\ and\ \citenamefont
  {Wen}}]{PhysRevB.84.214527}%
  \BibitemOpen
  \bibfield  {author} {\bibinfo {author} {\bibfnamefont {S.}~\bibnamefont
  {Li}}, \bibinfo {author} {\bibfnamefont {B.}~\bibnamefont {Zeng}}, \bibinfo
  {author} {\bibfnamefont {X.}~\bibnamefont {Wan}}, \bibinfo {author}
  {\bibfnamefont {J.}~\bibnamefont {Tao}}, \bibinfo {author} {\bibfnamefont
  {F.}~\bibnamefont {Han}}, \bibinfo {author} {\bibfnamefont {H.}~\bibnamefont
  {Yang}}, \bibinfo {author} {\bibfnamefont {Z.}~\bibnamefont {Wang}},\ and\
  \bibinfo {author} {\bibfnamefont {H.-H.}\ \bibnamefont {Wen}},\ }\bibfield
  {title} {\bibinfo {title} {Anomalous properties in the normal and
  superconducting states of laru${}_{3}$si${}_{2}$},\ }\href
  {https://doi.org/10.1103/PhysRevB.84.214527} {\bibfield  {journal} {\bibinfo
  {journal} {Phys. Rev. B}\ }\textbf {\bibinfo {volume} {84}},\ \bibinfo
  {pages} {214527} (\bibinfo {year} {2011})}\BibitemShut {NoStop}%
\bibitem [{\citenamefont {Li}\ \emph {et~al.}(2012)\citenamefont {Li},
  \citenamefont {Tao}, \citenamefont {Wan}, \citenamefont {Ding}, \citenamefont
  {Yang},\ and\ \citenamefont {Wen}}]{PhysRevB.86.024513}%
  \BibitemOpen
  \bibfield  {author} {\bibinfo {author} {\bibfnamefont {S.}~\bibnamefont
  {Li}}, \bibinfo {author} {\bibfnamefont {J.}~\bibnamefont {Tao}}, \bibinfo
  {author} {\bibfnamefont {X.}~\bibnamefont {Wan}}, \bibinfo {author}
  {\bibfnamefont {X.}~\bibnamefont {Ding}}, \bibinfo {author} {\bibfnamefont
  {H.}~\bibnamefont {Yang}},\ and\ \bibinfo {author} {\bibfnamefont {H.-H.}\
  \bibnamefont {Wen}},\ }\bibfield  {title} {\bibinfo {title} {Distinct
  behaviors of suppression to superconductivity in laru${}_{3}$si${}_{2}$
  induced by fe and co dopants},\ }\href
  {https://doi.org/10.1103/PhysRevB.86.024513} {\bibfield  {journal} {\bibinfo
  {journal} {Phys. Rev. B}\ }\textbf {\bibinfo {volume} {86}},\ \bibinfo
  {pages} {024513} (\bibinfo {year} {2012})}\BibitemShut {NoStop}%
\bibitem [{\citenamefont {Li}\ \emph {et~al.}(2016)\citenamefont {Li},
  \citenamefont {Li},\ and\ \citenamefont {Wen}}]{PhysRevB.94.094523}%
  \BibitemOpen
  \bibfield  {author} {\bibinfo {author} {\bibfnamefont {B.}~\bibnamefont
  {Li}}, \bibinfo {author} {\bibfnamefont {S.}~\bibnamefont {Li}},\ and\
  \bibinfo {author} {\bibfnamefont {H.-H.}\ \bibnamefont {Wen}},\ }\bibfield
  {title} {\bibinfo {title} {Chemical doping effect in the
  ${\mathbf{laru}}_{3}{\mathbf{si}}_{2}$ superconductor with a kagome
  lattice},\ }\href {https://doi.org/10.1103/PhysRevB.94.094523} {\bibfield
  {journal} {\bibinfo  {journal} {Phys. Rev. B}\ }\textbf {\bibinfo {volume}
  {94}},\ \bibinfo {pages} {094523} (\bibinfo {year} {2016})}\BibitemShut
  {NoStop}%
\bibitem [{\citenamefont {Mielke}\ \emph {et~al.}(2021)\citenamefont {Mielke},
  \citenamefont {Qin}, \citenamefont {Yin}, \citenamefont {Nakamura},
  \citenamefont {Das}, \citenamefont {Guo}, \citenamefont {Khasanov},
  \citenamefont {Chang}, \citenamefont {Wang}, \citenamefont {Jia},
  \citenamefont {Nakatsuji}, \citenamefont {Amato}, \citenamefont {Luetkens},
  \citenamefont {Xu}, \citenamefont {Hasan},\ and\ \citenamefont
  {Guguchia}}]{PhysRevMaterials.5.034803}%
  \BibitemOpen
  \bibfield  {author} {\bibinfo {author} {\bibfnamefont {C.}~\bibnamefont
  {Mielke}}, \bibinfo {author} {\bibfnamefont {Y.}~\bibnamefont {Qin}},
  \bibinfo {author} {\bibfnamefont {J.-X.}\ \bibnamefont {Yin}}, \bibinfo
  {author} {\bibfnamefont {H.}~\bibnamefont {Nakamura}}, \bibinfo {author}
  {\bibfnamefont {D.}~\bibnamefont {Das}}, \bibinfo {author} {\bibfnamefont
  {K.}~\bibnamefont {Guo}}, \bibinfo {author} {\bibfnamefont {R.}~\bibnamefont
  {Khasanov}}, \bibinfo {author} {\bibfnamefont {J.}~\bibnamefont {Chang}},
  \bibinfo {author} {\bibfnamefont {Z.~Q.}\ \bibnamefont {Wang}}, \bibinfo
  {author} {\bibfnamefont {S.}~\bibnamefont {Jia}}, \bibinfo {author}
  {\bibfnamefont {S.}~\bibnamefont {Nakatsuji}}, \bibinfo {author}
  {\bibfnamefont {A.}~\bibnamefont {Amato}}, \bibinfo {author} {\bibfnamefont
  {H.}~\bibnamefont {Luetkens}}, \bibinfo {author} {\bibfnamefont
  {G.}~\bibnamefont {Xu}}, \bibinfo {author} {\bibfnamefont {M.~Z.}\
  \bibnamefont {Hasan}},\ and\ \bibinfo {author} {\bibfnamefont
  {Z.}~\bibnamefont {Guguchia}},\ }\bibfield  {title} {\bibinfo {title}
  {Nodeless kagome superconductivity in
  ${\mathrm{laru}}_{3}{\mathrm{si}}_{2}$},\ }\href
  {https://doi.org/10.1103/PhysRevMaterials.5.034803} {\bibfield  {journal}
  {\bibinfo  {journal} {Phys. Rev. Mater.}\ }\textbf {\bibinfo {volume} {5}},\
  \bibinfo {pages} {034803} (\bibinfo {year} {2021})}\BibitemShut {NoStop}%
\bibitem [{\citenamefont {Liu}\ \emph {et~al.}(2024)\citenamefont {Liu},
  \citenamefont {Li}, \citenamefont {Yang}, \citenamefont {Lu}, \citenamefont
  {Cao}, \citenamefont {Li}, \citenamefont {Chai}, \citenamefont {Wu},
  \citenamefont {Li}, \citenamefont {Sun}, \citenamefont {Jiao}, \citenamefont
  {Wang}, \citenamefont {Xu}, \citenamefont {Ren},\ and\ \citenamefont
  {Cao}}]{Liu_2024}%
  \BibitemOpen
  \bibfield  {author} {\bibinfo {author} {\bibfnamefont {Y.}~\bibnamefont
  {Liu}}, \bibinfo {author} {\bibfnamefont {J.}~\bibnamefont {Li}}, \bibinfo
  {author} {\bibfnamefont {W.-Z.}\ \bibnamefont {Yang}}, \bibinfo {author}
  {\bibfnamefont {J.-Y.}\ \bibnamefont {Lu}}, \bibinfo {author} {\bibfnamefont
  {B.-Y.}\ \bibnamefont {Cao}}, \bibinfo {author} {\bibfnamefont {H.-X.}\
  \bibnamefont {Li}}, \bibinfo {author} {\bibfnamefont {W.-L.}\ \bibnamefont
  {Chai}}, \bibinfo {author} {\bibfnamefont {S.-Q.}\ \bibnamefont {Wu}},
  \bibinfo {author} {\bibfnamefont {B.-Z.}\ \bibnamefont {Li}}, \bibinfo
  {author} {\bibfnamefont {Y.-L.}\ \bibnamefont {Sun}}, \bibinfo {author}
  {\bibfnamefont {W.-H.}\ \bibnamefont {Jiao}}, \bibinfo {author}
  {\bibfnamefont {C.}~\bibnamefont {Wang}}, \bibinfo {author} {\bibfnamefont
  {X.-F.}\ \bibnamefont {Xu}}, \bibinfo {author} {\bibfnamefont
  {Z.}~\bibnamefont {Ren}},\ and\ \bibinfo {author} {\bibfnamefont {G.-H.}\
  \bibnamefont {Cao}},\ }\bibfield  {title} {\bibinfo {title}
  {Superconductivity in kagome metal thru3si2},\ }\href
  {https://doi.org/10.1088/1674-1056/ad1c5e} {\bibfield  {journal} {\bibinfo
  {journal} {Chinese Physics B}\ }\textbf {\bibinfo {volume} {33}},\ \bibinfo
  {pages} {057401} (\bibinfo {year} {2024})}\BibitemShut {NoStop}%
\bibitem [{\citenamefont {Chaudhary}\ \emph {et~al.}(2023)\citenamefont
  {Chaudhary}, \citenamefont {Shama}, \citenamefont {Singh}, \citenamefont
  {Consiglio}, \citenamefont {Di~Sante}, \citenamefont {Thomale},\ and\
  \citenamefont {Singh}}]{PhysRevB.107.085103}%
  \BibitemOpen
  \bibfield  {author} {\bibinfo {author} {\bibfnamefont {S.}~\bibnamefont
  {Chaudhary}}, \bibinfo {author} {\bibnamefont {Shama}}, \bibinfo {author}
  {\bibfnamefont {J.}~\bibnamefont {Singh}}, \bibinfo {author} {\bibfnamefont
  {A.}~\bibnamefont {Consiglio}}, \bibinfo {author} {\bibfnamefont
  {D.}~\bibnamefont {Di~Sante}}, \bibinfo {author} {\bibfnamefont
  {R.}~\bibnamefont {Thomale}},\ and\ \bibinfo {author} {\bibfnamefont
  {Y.}~\bibnamefont {Singh}},\ }\bibfield  {title} {\bibinfo {title} {Role of
  electronic correlations in the kagome-lattice superconductor
  ${\mathrm{larh}}_{3}{\mathrm{b}}_{2}$},\ }\href
  {https://doi.org/10.1103/PhysRevB.107.085103} {\bibfield  {journal} {\bibinfo
   {journal} {Phys. Rev. B}\ }\textbf {\bibinfo {volume} {107}},\ \bibinfo
  {pages} {085103} (\bibinfo {year} {2023})}\BibitemShut {NoStop}%
\bibitem [{\citenamefont {Kresse}\ and\ \citenamefont
  {Hafner}(1993)}]{PhysRevB.47.558}%
  \BibitemOpen
  \bibfield  {author} {\bibinfo {author} {\bibfnamefont {G.}~\bibnamefont
  {Kresse}}\ and\ \bibinfo {author} {\bibfnamefont {J.}~\bibnamefont
  {Hafner}},\ }\bibfield  {title} {\bibinfo {title} {Ab initio molecular
  dynamics for liquid metals},\ }\href
  {https://doi.org/10.1103/PhysRevB.47.558} {\bibfield  {journal} {\bibinfo
  {journal} {Phys. Rev. B}\ }\textbf {\bibinfo {volume} {47}},\ \bibinfo
  {pages} {558} (\bibinfo {year} {1993})}\BibitemShut {NoStop}%
\bibitem [{\citenamefont {Kresse}\ and\ \citenamefont
  {Hafner}(1994)}]{PhysRevB.49.14251}%
  \BibitemOpen
  \bibfield  {author} {\bibinfo {author} {\bibfnamefont {G.}~\bibnamefont
  {Kresse}}\ and\ \bibinfo {author} {\bibfnamefont {J.}~\bibnamefont
  {Hafner}},\ }\bibfield  {title} {\bibinfo {title} {Ab initio
  molecular-dynamics simulation of the liquid-metal--amorphous-semiconductor
  transition in germanium},\ }\href {https://doi.org/10.1103/PhysRevB.49.14251}
  {\bibfield  {journal} {\bibinfo  {journal} {Phys. Rev. B}\ }\textbf {\bibinfo
  {volume} {49}},\ \bibinfo {pages} {14251} (\bibinfo {year}
  {1994})}\BibitemShut {NoStop}%
\bibitem [{\citenamefont {Kresse}\ and\ \citenamefont
  {Furthm\"uller}(1996)}]{PhysRevB.54.11169}%
  \BibitemOpen
  \bibfield  {author} {\bibinfo {author} {\bibfnamefont {G.}~\bibnamefont
  {Kresse}}\ and\ \bibinfo {author} {\bibfnamefont {J.}~\bibnamefont
  {Furthm\"uller}},\ }\bibfield  {title} {\bibinfo {title} {Efficient iterative
  schemes for ab initio total-energy calculations using a plane-wave basis
  set},\ }\href {https://doi.org/10.1103/PhysRevB.54.11169} {\bibfield
  {journal} {\bibinfo  {journal} {Phys. Rev. B}\ }\textbf {\bibinfo {volume}
  {54}},\ \bibinfo {pages} {11169} (\bibinfo {year} {1996})}\BibitemShut
  {NoStop}%
\bibitem [{\citenamefont {Kresse}\ and\ \citenamefont
  {Furthmüller}(1996)}]{KRESSE199615}%
  \BibitemOpen
  \bibfield  {author} {\bibinfo {author} {\bibfnamefont {G.}~\bibnamefont
  {Kresse}}\ and\ \bibinfo {author} {\bibfnamefont {J.}~\bibnamefont
  {Furthmüller}},\ }\bibfield  {title} {\bibinfo {title} {Efficiency of
  ab-initio total energy calculations for metals and semiconductors using a
  plane-wave basis set},\ }\href
  {https://doi.org/https://doi.org/10.1016/0927-0256(96)00008-0} {\bibfield
  {journal} {\bibinfo  {journal} {Computational Materials Science}\ }\textbf
  {\bibinfo {volume} {6}},\ \bibinfo {pages} {15} (\bibinfo {year}
  {1996})}\BibitemShut {NoStop}%
\bibitem [{\citenamefont {Kresse}\ and\ \citenamefont
  {Joubert}(1999)}]{PhysRevB.59.1758}%
  \BibitemOpen
  \bibfield  {author} {\bibinfo {author} {\bibfnamefont {G.}~\bibnamefont
  {Kresse}}\ and\ \bibinfo {author} {\bibfnamefont {D.}~\bibnamefont
  {Joubert}},\ }\bibfield  {title} {\bibinfo {title} {From ultrasoft
  pseudopotentials to the projector augmented-wave method},\ }\href
  {https://doi.org/10.1103/PhysRevB.59.1758} {\bibfield  {journal} {\bibinfo
  {journal} {Phys. Rev. B}\ }\textbf {\bibinfo {volume} {59}},\ \bibinfo
  {pages} {1758} (\bibinfo {year} {1999})}\BibitemShut {NoStop}%
\bibitem [{\citenamefont {Perdew}\ \emph {et~al.}(1996)\citenamefont {Perdew},
  \citenamefont {Burke},\ and\ \citenamefont
  {Ernzerhof}}]{PhysRevLett.77.3865}%
  \BibitemOpen
  \bibfield  {author} {\bibinfo {author} {\bibfnamefont {J.~P.}\ \bibnamefont
  {Perdew}}, \bibinfo {author} {\bibfnamefont {K.}~\bibnamefont {Burke}},\ and\
  \bibinfo {author} {\bibfnamefont {M.}~\bibnamefont {Ernzerhof}},\ }\bibfield
  {title} {\bibinfo {title} {Generalized gradient approximation made simple},\
  }\href {https://doi.org/10.1103/PhysRevLett.77.3865} {\bibfield  {journal}
  {\bibinfo  {journal} {Phys. Rev. Lett.}\ }\textbf {\bibinfo {volume} {77}},\
  \bibinfo {pages} {3865} (\bibinfo {year} {1996})}\BibitemShut {NoStop}%
\bibitem [{\citenamefont {Giannozzi}\ \emph {et~al.}(2020)\citenamefont
  {Giannozzi}, \citenamefont {Baseggio}, \citenamefont {Bonfà}, \citenamefont
  {Brunato}, \citenamefont {Car}, \citenamefont {Carnimeo}, \citenamefont
  {Cavazzoni}, \citenamefont {de~Gironcoli}, \citenamefont {Delugas},
  \citenamefont {Ferrari~Ruffino}, \citenamefont {Ferretti}, \citenamefont
  {Marzari}, \citenamefont {Timrov}, \citenamefont {Urru},\ and\ \citenamefont
  {Baroni}}]{10.1063/5.0005082}%
  \BibitemOpen
  \bibfield  {author} {\bibinfo {author} {\bibfnamefont {P.}~\bibnamefont
  {Giannozzi}}, \bibinfo {author} {\bibfnamefont {O.}~\bibnamefont {Baseggio}},
  \bibinfo {author} {\bibfnamefont {P.}~\bibnamefont {Bonfà}}, \bibinfo
  {author} {\bibfnamefont {D.}~\bibnamefont {Brunato}}, \bibinfo {author}
  {\bibfnamefont {R.}~\bibnamefont {Car}}, \bibinfo {author} {\bibfnamefont
  {I.}~\bibnamefont {Carnimeo}}, \bibinfo {author} {\bibfnamefont
  {C.}~\bibnamefont {Cavazzoni}}, \bibinfo {author} {\bibfnamefont
  {S.}~\bibnamefont {de~Gironcoli}}, \bibinfo {author} {\bibfnamefont
  {P.}~\bibnamefont {Delugas}}, \bibinfo {author} {\bibfnamefont
  {F.}~\bibnamefont {Ferrari~Ruffino}}, \bibinfo {author} {\bibfnamefont
  {A.}~\bibnamefont {Ferretti}}, \bibinfo {author} {\bibfnamefont
  {N.}~\bibnamefont {Marzari}}, \bibinfo {author} {\bibfnamefont
  {I.}~\bibnamefont {Timrov}}, \bibinfo {author} {\bibfnamefont
  {A.}~\bibnamefont {Urru}},\ and\ \bibinfo {author} {\bibfnamefont
  {S.}~\bibnamefont {Baroni}},\ }\bibfield  {title} {\bibinfo {title} {Quantum
  espresso toward the exascale},\ }\href {https://doi.org/10.1063/5.0005082}
  {\bibfield  {journal} {\bibinfo  {journal} {The Journal of Chemical Physics}\
  }\textbf {\bibinfo {volume} {152}},\ \bibinfo {pages} {154105} (\bibinfo
  {year} {2020})},\ \Eprint
  {https://arxiv.org/abs/https://pubs.aip.org/aip/jcp/article-pdf/doi/10.1063/5.0005082/16721881/154105\_1\_online.pdf}
  {https://pubs.aip.org/aip/jcp/article-pdf/doi/10.1063/5.0005082/16721881/154105\_1\_online.pdf}
  \BibitemShut {NoStop}%
\bibitem [{\citenamefont {Giannozzi}\ \emph {et~al.}(2009)\citenamefont
  {Giannozzi}, \citenamefont {Baroni}, \citenamefont {Bonini}, \citenamefont
  {Calandra}, \citenamefont {Car}, \citenamefont {Cavazzoni}, \citenamefont
  {Ceresoli}, \citenamefont {Chiarotti}, \citenamefont {Cococcioni},
  \citenamefont {Dabo}, \citenamefont {Dal~Corso}, \citenamefont
  {de~Gironcoli}, \citenamefont {Fabris}, \citenamefont {Fratesi},
  \citenamefont {Gebauer}, \citenamefont {Gerstmann}, \citenamefont
  {Gougoussis}, \citenamefont {Kokalj}, \citenamefont {Lazzeri}, \citenamefont
  {Martin-Samos}, \citenamefont {Marzari}, \citenamefont {Mauri}, \citenamefont
  {Mazzarello}, \citenamefont {Paolini}, \citenamefont {Pasquarello},
  \citenamefont {Paulatto}, \citenamefont {Sbraccia}, \citenamefont {Scandolo},
  \citenamefont {Sclauzero}, \citenamefont {Seitsonen}, \citenamefont
  {Smogunov}, \citenamefont {Umari},\ and\ \citenamefont
  {Wentzcovitch}}]{Giannozzi_2009}%
  \BibitemOpen
  \bibfield  {author} {\bibinfo {author} {\bibfnamefont {P.}~\bibnamefont
  {Giannozzi}}, \bibinfo {author} {\bibfnamefont {S.}~\bibnamefont {Baroni}},
  \bibinfo {author} {\bibfnamefont {N.}~\bibnamefont {Bonini}}, \bibinfo
  {author} {\bibfnamefont {M.}~\bibnamefont {Calandra}}, \bibinfo {author}
  {\bibfnamefont {R.}~\bibnamefont {Car}}, \bibinfo {author} {\bibfnamefont
  {C.}~\bibnamefont {Cavazzoni}}, \bibinfo {author} {\bibfnamefont
  {D.}~\bibnamefont {Ceresoli}}, \bibinfo {author} {\bibfnamefont {G.~L.}\
  \bibnamefont {Chiarotti}}, \bibinfo {author} {\bibfnamefont {M.}~\bibnamefont
  {Cococcioni}}, \bibinfo {author} {\bibfnamefont {I.}~\bibnamefont {Dabo}},
  \bibinfo {author} {\bibfnamefont {A.}~\bibnamefont {Dal~Corso}}, \bibinfo
  {author} {\bibfnamefont {S.}~\bibnamefont {de~Gironcoli}}, \bibinfo {author}
  {\bibfnamefont {S.}~\bibnamefont {Fabris}}, \bibinfo {author} {\bibfnamefont
  {G.}~\bibnamefont {Fratesi}}, \bibinfo {author} {\bibfnamefont
  {R.}~\bibnamefont {Gebauer}}, \bibinfo {author} {\bibfnamefont
  {U.}~\bibnamefont {Gerstmann}}, \bibinfo {author} {\bibfnamefont
  {C.}~\bibnamefont {Gougoussis}}, \bibinfo {author} {\bibfnamefont
  {A.}~\bibnamefont {Kokalj}}, \bibinfo {author} {\bibfnamefont
  {M.}~\bibnamefont {Lazzeri}}, \bibinfo {author} {\bibfnamefont
  {L.}~\bibnamefont {Martin-Samos}}, \bibinfo {author} {\bibfnamefont
  {N.}~\bibnamefont {Marzari}}, \bibinfo {author} {\bibfnamefont
  {F.}~\bibnamefont {Mauri}}, \bibinfo {author} {\bibfnamefont
  {R.}~\bibnamefont {Mazzarello}}, \bibinfo {author} {\bibfnamefont
  {S.}~\bibnamefont {Paolini}}, \bibinfo {author} {\bibfnamefont
  {A.}~\bibnamefont {Pasquarello}}, \bibinfo {author} {\bibfnamefont
  {L.}~\bibnamefont {Paulatto}}, \bibinfo {author} {\bibfnamefont
  {C.}~\bibnamefont {Sbraccia}}, \bibinfo {author} {\bibfnamefont
  {S.}~\bibnamefont {Scandolo}}, \bibinfo {author} {\bibfnamefont
  {G.}~\bibnamefont {Sclauzero}}, \bibinfo {author} {\bibfnamefont {A.~P.}\
  \bibnamefont {Seitsonen}}, \bibinfo {author} {\bibfnamefont {A.}~\bibnamefont
  {Smogunov}}, \bibinfo {author} {\bibfnamefont {P.}~\bibnamefont {Umari}},\
  and\ \bibinfo {author} {\bibfnamefont {R.~M.}\ \bibnamefont {Wentzcovitch}},\
  }\bibfield  {title} {\bibinfo {title} {Quantum espresso: a modular and
  open-source software project for quantum simulations of materials},\ }\href
  {https://doi.org/10.1088/0953-8984/21/39/395502} {\bibfield  {journal}
  {\bibinfo  {journal} {Journal of Physics: Condensed Matter}\ }\textbf
  {\bibinfo {volume} {21}},\ \bibinfo {pages} {395502} (\bibinfo {year}
  {2009})}\BibitemShut {NoStop}%
\bibitem [{\citenamefont {Giannozzi}\ \emph {et~al.}(2017)\citenamefont
  {Giannozzi}, \citenamefont {Andreussi}, \citenamefont {Brumme}, \citenamefont
  {Bunau}, \citenamefont {Buongiorno~Nardelli}, \citenamefont {Calandra},
  \citenamefont {Car}, \citenamefont {Cavazzoni}, \citenamefont {Ceresoli},
  \citenamefont {Cococcioni}, \citenamefont {Colonna}, \citenamefont
  {Carnimeo}, \citenamefont {Dal~Corso}, \citenamefont {de~Gironcoli},
  \citenamefont {Delugas}, \citenamefont {DiStasio}, \citenamefont {Ferretti},
  \citenamefont {Floris}, \citenamefont {Fratesi}, \citenamefont {Fugallo},
  \citenamefont {Gebauer}, \citenamefont {Gerstmann}, \citenamefont {Giustino},
  \citenamefont {Gorni}, \citenamefont {Jia}, \citenamefont {Kawamura},
  \citenamefont {Ko}, \citenamefont {Kokalj}, \citenamefont {Küçükbenli},
  \citenamefont {Lazzeri}, \citenamefont {Marsili}, \citenamefont {Marzari},
  \citenamefont {Mauri}, \citenamefont {Nguyen}, \citenamefont {Nguyen},
  \citenamefont {Otero-de-la Roza}, \citenamefont {Paulatto}, \citenamefont
  {Poncé}, \citenamefont {Rocca}, \citenamefont {Sabatini}, \citenamefont
  {Santra}, \citenamefont {Schlipf}, \citenamefont {Seitsonen}, \citenamefont
  {Smogunov}, \citenamefont {Timrov}, \citenamefont {Thonhauser}, \citenamefont
  {Umari}, \citenamefont {Vast}, \citenamefont {Wu},\ and\ \citenamefont
  {Baroni}}]{Giannozzi_2017}%
  \BibitemOpen
  \bibfield  {author} {\bibinfo {author} {\bibfnamefont {P.}~\bibnamefont
  {Giannozzi}}, \bibinfo {author} {\bibfnamefont {O.}~\bibnamefont
  {Andreussi}}, \bibinfo {author} {\bibfnamefont {T.}~\bibnamefont {Brumme}},
  \bibinfo {author} {\bibfnamefont {O.}~\bibnamefont {Bunau}}, \bibinfo
  {author} {\bibfnamefont {M.}~\bibnamefont {Buongiorno~Nardelli}}, \bibinfo
  {author} {\bibfnamefont {M.}~\bibnamefont {Calandra}}, \bibinfo {author}
  {\bibfnamefont {R.}~\bibnamefont {Car}}, \bibinfo {author} {\bibfnamefont
  {C.}~\bibnamefont {Cavazzoni}}, \bibinfo {author} {\bibfnamefont
  {D.}~\bibnamefont {Ceresoli}}, \bibinfo {author} {\bibfnamefont
  {M.}~\bibnamefont {Cococcioni}}, \bibinfo {author} {\bibfnamefont
  {N.}~\bibnamefont {Colonna}}, \bibinfo {author} {\bibfnamefont
  {I.}~\bibnamefont {Carnimeo}}, \bibinfo {author} {\bibfnamefont
  {A.}~\bibnamefont {Dal~Corso}}, \bibinfo {author} {\bibfnamefont
  {S.}~\bibnamefont {de~Gironcoli}}, \bibinfo {author} {\bibfnamefont
  {P.}~\bibnamefont {Delugas}}, \bibinfo {author} {\bibfnamefont {R.~A.}\
  \bibnamefont {DiStasio}}, \bibinfo {author} {\bibfnamefont {A.}~\bibnamefont
  {Ferretti}}, \bibinfo {author} {\bibfnamefont {A.}~\bibnamefont {Floris}},
  \bibinfo {author} {\bibfnamefont {G.}~\bibnamefont {Fratesi}}, \bibinfo
  {author} {\bibfnamefont {G.}~\bibnamefont {Fugallo}}, \bibinfo {author}
  {\bibfnamefont {R.}~\bibnamefont {Gebauer}}, \bibinfo {author} {\bibfnamefont
  {U.}~\bibnamefont {Gerstmann}}, \bibinfo {author} {\bibfnamefont
  {F.}~\bibnamefont {Giustino}}, \bibinfo {author} {\bibfnamefont
  {T.}~\bibnamefont {Gorni}}, \bibinfo {author} {\bibfnamefont
  {J.}~\bibnamefont {Jia}}, \bibinfo {author} {\bibfnamefont {M.}~\bibnamefont
  {Kawamura}}, \bibinfo {author} {\bibfnamefont {H.-Y.}\ \bibnamefont {Ko}},
  \bibinfo {author} {\bibfnamefont {A.}~\bibnamefont {Kokalj}}, \bibinfo
  {author} {\bibfnamefont {E.}~\bibnamefont {Küçükbenli}}, \bibinfo {author}
  {\bibfnamefont {M.}~\bibnamefont {Lazzeri}}, \bibinfo {author} {\bibfnamefont
  {M.}~\bibnamefont {Marsili}}, \bibinfo {author} {\bibfnamefont
  {N.}~\bibnamefont {Marzari}}, \bibinfo {author} {\bibfnamefont
  {F.}~\bibnamefont {Mauri}}, \bibinfo {author} {\bibfnamefont {N.~L.}\
  \bibnamefont {Nguyen}}, \bibinfo {author} {\bibfnamefont {H.-V.}\
  \bibnamefont {Nguyen}}, \bibinfo {author} {\bibfnamefont {A.}~\bibnamefont
  {Otero-de-la Roza}}, \bibinfo {author} {\bibfnamefont {L.}~\bibnamefont
  {Paulatto}}, \bibinfo {author} {\bibfnamefont {S.}~\bibnamefont {Poncé}},
  \bibinfo {author} {\bibfnamefont {D.}~\bibnamefont {Rocca}}, \bibinfo
  {author} {\bibfnamefont {R.}~\bibnamefont {Sabatini}}, \bibinfo {author}
  {\bibfnamefont {B.}~\bibnamefont {Santra}}, \bibinfo {author} {\bibfnamefont
  {M.}~\bibnamefont {Schlipf}}, \bibinfo {author} {\bibfnamefont {A.~P.}\
  \bibnamefont {Seitsonen}}, \bibinfo {author} {\bibfnamefont {A.}~\bibnamefont
  {Smogunov}}, \bibinfo {author} {\bibfnamefont {I.}~\bibnamefont {Timrov}},
  \bibinfo {author} {\bibfnamefont {T.}~\bibnamefont {Thonhauser}}, \bibinfo
  {author} {\bibfnamefont {P.}~\bibnamefont {Umari}}, \bibinfo {author}
  {\bibfnamefont {N.}~\bibnamefont {Vast}}, \bibinfo {author} {\bibfnamefont
  {X.}~\bibnamefont {Wu}},\ and\ \bibinfo {author} {\bibfnamefont
  {S.}~\bibnamefont {Baroni}},\ }\bibfield  {title} {\bibinfo {title} {Advanced
  capabilities for materials modelling with quantum espresso},\ }\href
  {https://doi.org/10.1088/1361-648X/aa8f79} {\bibfield  {journal} {\bibinfo
  {journal} {Journal of Physics: Condensed Matter}\ }\textbf {\bibinfo {volume}
  {29}},\ \bibinfo {pages} {465901} (\bibinfo {year} {2017})}\BibitemShut
  {NoStop}%
\bibitem [{\citenamefont {Hamann}(2013)}]{PhysRevB.88.085117}%
  \BibitemOpen
  \bibfield  {author} {\bibinfo {author} {\bibfnamefont {D.~R.}\ \bibnamefont
  {Hamann}},\ }\bibfield  {title} {\bibinfo {title} {Optimized norm-conserving
  vanderbilt pseudopotentials},\ }\href
  {https://doi.org/10.1103/PhysRevB.88.085117} {\bibfield  {journal} {\bibinfo
  {journal} {Phys. Rev. B}\ }\textbf {\bibinfo {volume} {88}},\ \bibinfo
  {pages} {085117} (\bibinfo {year} {2013})}\BibitemShut {NoStop}%
\bibitem [{\citenamefont {Wierzbowska}\ \emph {et~al.}(2006)\citenamefont
  {Wierzbowska}, \citenamefont {de~Gironcoli},\ and\ \citenamefont
  {Giannozzi}}]{wierzbowska2006originslowhighpressurediscontinuities}%
  \BibitemOpen
  \bibfield  {author} {\bibinfo {author} {\bibfnamefont {M.}~\bibnamefont
  {Wierzbowska}}, \bibinfo {author} {\bibfnamefont {S.}~\bibnamefont
  {de~Gironcoli}},\ and\ \bibinfo {author} {\bibfnamefont {P.}~\bibnamefont
  {Giannozzi}},\ }\href {https://arxiv.org/abs/cond-mat/0504077} {\bibinfo
  {title} {Origins of low- and high-pressure discontinuities of $t_{c}$ in
  niobium}} (\bibinfo {year} {2006}),\ \Eprint
  {https://arxiv.org/abs/cond-mat/0504077} {arXiv:cond-mat/0504077
  [cond-mat.supr-con]} \BibitemShut {NoStop}%
\bibitem [{\citenamefont {Chacon}\ and\ \citenamefont
  {Isnard}(2001)}]{Chacon2001}%
  \BibitemOpen
  \bibfield  {author} {\bibinfo {author} {\bibfnamefont {C.}~\bibnamefont
  {Chacon}}\ and\ \bibinfo {author} {\bibfnamefont {O.}~\bibnamefont
  {Isnard}},\ }\bibfield  {title} {\bibinfo {title} {The structural and
  magnetic properties of yn+1co3n+5b2n compounds investigated by neutron
  diffraction},\ }\href {https://doi.org/10.1088/0953-8984/13/25/310}
  {\bibfield  {journal} {\bibinfo  {journal} {Journal of Physics: Condensed
  Matter}\ }\textbf {\bibinfo {volume} {13}},\ \bibinfo {pages} {5841}
  (\bibinfo {year} {2001})}\BibitemShut {NoStop}%
\bibitem [{\citenamefont {Mendelsohn}\ \emph {et~al.}(1970)\citenamefont
  {Mendelsohn}, \citenamefont {Biggs},\ and\ \citenamefont
  {Mann}}]{PhysRevA.2.1130}%
  \BibitemOpen
  \bibfield  {author} {\bibinfo {author} {\bibfnamefont {L.~B.}\ \bibnamefont
  {Mendelsohn}}, \bibinfo {author} {\bibfnamefont {F.}~\bibnamefont {Biggs}},\
  and\ \bibinfo {author} {\bibfnamefont {J.~B.}\ \bibnamefont {Mann}},\
  }\bibfield  {title} {\bibinfo {title} {Hartree-fock diamagnetic
  susceptibilities},\ }\href {https://doi.org/10.1103/PhysRevA.2.1130}
  {\bibfield  {journal} {\bibinfo  {journal} {Phys. Rev. A}\ }\textbf {\bibinfo
  {volume} {2}},\ \bibinfo {pages} {1130} (\bibinfo {year} {1970})}\BibitemShut
  {NoStop}%
\bibitem [{\citenamefont {Singh}\ \emph {et~al.}(2007)\citenamefont {Singh},
  \citenamefont {Niazi}, \citenamefont {Vannette}, \citenamefont {Prozorov},\
  and\ \citenamefont {Johnston}}]{PhysRevB.76.214510}%
  \BibitemOpen
  \bibfield  {author} {\bibinfo {author} {\bibfnamefont {Y.}~\bibnamefont
  {Singh}}, \bibinfo {author} {\bibfnamefont {A.}~\bibnamefont {Niazi}},
  \bibinfo {author} {\bibfnamefont {M.~D.}\ \bibnamefont {Vannette}}, \bibinfo
  {author} {\bibfnamefont {R.}~\bibnamefont {Prozorov}},\ and\ \bibinfo
  {author} {\bibfnamefont {D.~C.}\ \bibnamefont {Johnston}},\ }\bibfield
  {title} {\bibinfo {title} {Superconducting and normal-state properties of the
  layered boride $\mathrm{Os}{\mathrm{b}}_{2}$},\ }\href
  {https://doi.org/10.1103/PhysRevB.76.214510} {\bibfield  {journal} {\bibinfo
  {journal} {Phys. Rev. B}\ }\textbf {\bibinfo {volume} {76}},\ \bibinfo
  {pages} {214510} (\bibinfo {year} {2007})}\BibitemShut {NoStop}%
\bibitem [{\citenamefont {Wilson}(1975)}]{RevModPhys.47.773}%
  \BibitemOpen
  \bibfield  {author} {\bibinfo {author} {\bibfnamefont {K.~G.}\ \bibnamefont
  {Wilson}},\ }\bibfield  {title} {\bibinfo {title} {The renormalization group:
  Critical phenomena and the kondo problem},\ }\href
  {https://doi.org/10.1103/RevModPhys.47.773} {\bibfield  {journal} {\bibinfo
  {journal} {Rev. Mod. Phys.}\ }\textbf {\bibinfo {volume} {47}},\ \bibinfo
  {pages} {773} (\bibinfo {year} {1975})}\BibitemShut {NoStop}%
\bibitem [{\citenamefont {Gong}\ \emph {et~al.}(2022)\citenamefont {Gong},
  \citenamefont {Tian}, \citenamefont {Tu}, \citenamefont {Yin}, \citenamefont
  {Fu}, \citenamefont {Luo},\ and\ \citenamefont {Lei}}]{Gong_2022}%
  \BibitemOpen
  \bibfield  {author} {\bibinfo {author} {\bibfnamefont {C.}~\bibnamefont
  {Gong}}, \bibinfo {author} {\bibfnamefont {S.}~\bibnamefont {Tian}}, \bibinfo
  {author} {\bibfnamefont {Z.}~\bibnamefont {Tu}}, \bibinfo {author}
  {\bibfnamefont {Q.}~\bibnamefont {Yin}}, \bibinfo {author} {\bibfnamefont
  {Y.}~\bibnamefont {Fu}}, \bibinfo {author} {\bibfnamefont {R.}~\bibnamefont
  {Luo}},\ and\ \bibinfo {author} {\bibfnamefont {H.}~\bibnamefont {Lei}},\
  }\bibfield  {title} {\bibinfo {title} {Superconductivity in kagome metal
  yru3si2 with strong electron correlations},\ }\href
  {https://doi.org/10.1088/0256-307X/39/8/087401} {\bibfield  {journal}
  {\bibinfo  {journal} {Chinese Physics Letters}\ }\textbf {\bibinfo {volume}
  {39}},\ \bibinfo {pages} {087401} (\bibinfo {year} {2022})}\BibitemShut
  {NoStop}%
\bibitem [{\citenamefont {Rice}(1968)}]{PhysRevLett.20.1439}%
  \BibitemOpen
  \bibfield  {author} {\bibinfo {author} {\bibfnamefont {M.~J.}\ \bibnamefont
  {Rice}},\ }\bibfield  {title} {\bibinfo {title} {Electron-electron scattering
  in transition metals},\ }\href {https://doi.org/10.1103/PhysRevLett.20.1439}
  {\bibfield  {journal} {\bibinfo  {journal} {Phys. Rev. Lett.}\ }\textbf
  {\bibinfo {volume} {20}},\ \bibinfo {pages} {1439} (\bibinfo {year}
  {1968})}\BibitemShut {NoStop}%
\bibitem [{\citenamefont {Kadowaki}\ and\ \citenamefont
  {Woods}(1986)}]{KADOWAKI1986507}%
  \BibitemOpen
  \bibfield  {author} {\bibinfo {author} {\bibfnamefont {K.}~\bibnamefont
  {Kadowaki}}\ and\ \bibinfo {author} {\bibfnamefont {S.}~\bibnamefont
  {Woods}},\ }\bibfield  {title} {\bibinfo {title} {Universal relationship of
  the resistivity and specific heat in heavy-fermion compounds},\ }\href
  {https://doi.org/https://doi.org/10.1016/0038-1098(86)90785-4} {\bibfield
  {journal} {\bibinfo  {journal} {Solid State Communications}\ }\textbf
  {\bibinfo {volume} {58}},\ \bibinfo {pages} {507} (\bibinfo {year}
  {1986})}\BibitemShut {NoStop}%
\bibitem [{\citenamefont {Singh}\ \emph {et~al.}(2010)\citenamefont {Singh},
  \citenamefont {Martin}, \citenamefont {Bud'ko}, \citenamefont {Ellern},
  \citenamefont {Prozorov},\ and\ \citenamefont
  {Johnston}}]{PhysRevB.82.144532}%
  \BibitemOpen
  \bibfield  {author} {\bibinfo {author} {\bibfnamefont {Y.}~\bibnamefont
  {Singh}}, \bibinfo {author} {\bibfnamefont {C.}~\bibnamefont {Martin}},
  \bibinfo {author} {\bibfnamefont {S.~L.}\ \bibnamefont {Bud'ko}}, \bibinfo
  {author} {\bibfnamefont {A.}~\bibnamefont {Ellern}}, \bibinfo {author}
  {\bibfnamefont {R.}~\bibnamefont {Prozorov}},\ and\ \bibinfo {author}
  {\bibfnamefont {D.~C.}\ \bibnamefont {Johnston}},\ }\bibfield  {title}
  {\bibinfo {title} {Multigap superconductivity and shubnikov--de haas
  oscillations in single crystals of the layered boride ${\text{osb}}_{2}$},\
  }\href {https://doi.org/10.1103/PhysRevB.82.144532} {\bibfield  {journal}
  {\bibinfo  {journal} {Phys. Rev. B}\ }\textbf {\bibinfo {volume} {82}},\
  \bibinfo {pages} {144532} (\bibinfo {year} {2010})}\BibitemShut {NoStop}%
\bibitem [{\citenamefont {McMillan}(1968)}]{PhysRev.167.331}%
  \BibitemOpen
  \bibfield  {author} {\bibinfo {author} {\bibfnamefont {W.~L.}\ \bibnamefont
  {McMillan}},\ }\bibfield  {title} {\bibinfo {title} {Transition temperature
  of strong-coupled superconductors},\ }\href
  {https://doi.org/10.1103/PhysRev.167.331} {\bibfield  {journal} {\bibinfo
  {journal} {Phys. Rev.}\ }\textbf {\bibinfo {volume} {167}},\ \bibinfo {pages}
  {331} (\bibinfo {year} {1968})}\BibitemShut {NoStop}%
\bibitem [{\citenamefont {Carbotte}(1990)}]{RevModPhys.62.1027}%
  \BibitemOpen
  \bibfield  {author} {\bibinfo {author} {\bibfnamefont {J.~P.}\ \bibnamefont
  {Carbotte}},\ }\bibfield  {title} {\bibinfo {title} {Properties of
  boson-exchange superconductors},\ }\href
  {https://doi.org/10.1103/RevModPhys.62.1027} {\bibfield  {journal} {\bibinfo
  {journal} {Rev. Mod. Phys.}\ }\textbf {\bibinfo {volume} {62}},\ \bibinfo
  {pages} {1027} (\bibinfo {year} {1990})}\BibitemShut {NoStop}%
\bibitem [{\citenamefont {Xiao}\ \emph {et~al.}(2025)\citenamefont {Xiao},
  \citenamefont {Duan}, \citenamefont {jia}, \citenamefont {Cui}, \citenamefont
  {Liu}, \citenamefont {Wen}, \citenamefont {Ji}, \citenamefont {Zhong},
  \citenamefont {Chen},\ and\ \citenamefont {Zhao}}]{PhysRevB.111.195132}%
  \BibitemOpen
  \bibfield  {author} {\bibinfo {author} {\bibfnamefont {Y.}~\bibnamefont
  {Xiao}}, \bibinfo {author} {\bibfnamefont {Q.}~\bibnamefont {Duan}}, \bibinfo
  {author} {\bibfnamefont {T.}~\bibnamefont {jia}}, \bibinfo {author}
  {\bibfnamefont {Y.}~\bibnamefont {Cui}}, \bibinfo {author} {\bibfnamefont
  {S.}~\bibnamefont {Liu}}, \bibinfo {author} {\bibfnamefont {Z.}~\bibnamefont
  {Wen}}, \bibinfo {author} {\bibfnamefont {L.}~\bibnamefont {Ji}}, \bibinfo
  {author} {\bibfnamefont {R.}~\bibnamefont {Zhong}}, \bibinfo {author}
  {\bibfnamefont {Y.}~\bibnamefont {Chen}},\ and\ \bibinfo {author}
  {\bibfnamefont {Y.}~\bibnamefont {Zhao}},\ }\bibfield  {title} {\bibinfo
  {title} {Superconductivity and electron correlations in the kagome metal
  ${\text{luos}}_{3}{\text{b}}_{2}$},\ }\href
  {https://doi.org/10.1103/PhysRevB.111.195132} {\bibfield  {journal} {\bibinfo
   {journal} {Phys. Rev. B}\ }\textbf {\bibinfo {volume} {111}},\ \bibinfo
  {pages} {195132} (\bibinfo {year} {2025})}\BibitemShut {NoStop}%
\end{thebibliography}%

\end{document}